\documentclass[reprint,prb,aps,longbibliography,nofootinbib,floatfix,nobalancelastpage]{revtex4-2}
\usepackage{amsmath}
\usepackage{amsfonts}
\usepackage{amssymb}
\usepackage{mathtools}
\usepackage{graphicx}
\usepackage[dvipsnames]{xcolor}
\usepackage{yfonts}
\usepackage{bm}
\usepackage{ulem}
\usepackage{dsfont}
\usepackage{braket}
\usepackage{mathrsfs}

\definecolor{myBlue}{RGB}{31,119,180}
\definecolor{myOrange}{RGB}{255,127,14}
\definecolor{myGreen}{RGB}{44,160,44}
\definecolor{myRed}{RGB}{214,39,40}
\definecolor{myPurple}{RGB}{148,103,189}

\usepackage[colorlinks=True,citecolor=gray,linkcolor=myRed,urlcolor=gray]{hyperref}
\usepackage{hypcap}

\newcommand\abs[1]{|#1|}
\newcommand\eq[1]{\begin{align}#1\end{align}}
\newcommand{\eps}{\epsilon}
\newcommand{\levy}{{\rm L{\acute{e}}vy}}

\newcommand\new[1]{{#1}}

\begin{document}

\title{Anderson localisation in spatially structured random graphs}

\author{Bibek Saha}
\email{bibek.saha@icts.res.in}
\affiliation{International Centre for Theoretical Sciences, Tata Institute of Fundamental Research, Bengaluru 560089, India}

\author{Sthitadhi Roy}
\email{sthitadhi.roy@icts.res.in}
\affiliation{International Centre for Theoretical Sciences, Tata Institute of Fundamental Research, Bengaluru 560089, India}

\date{\today}
\begin{abstract}
We study Anderson localisation on high-dimensional graphs with spatial structure induced by long-ranged but distance-dependent hopping. 
To this end, we introduce a class of models that interpolate between the short-range Anderson model on a random regular graph and fully connected models with statistically uniform hopping, by embedding a random regular graph into a complete graph and allowing hopping amplitudes to decay exponentially with graph distance.
The competition between the exponentially growing number of neighbours with graph distance and the exponentially decaying hopping amplitude positions our models effectively as power-law hopping generalisation of the Anderson model on random regular graphs.
Using a combination of numerical exact diagonalisation and analytical renormalised perturbation theory, we establish the resulting localisation phase diagram emerging from the interplay of the lengthscale associated to the hopping range and the onsite disorder strength. We find that increasing the hopping range shifts the localisation transition to stronger disorder, and that beyond a critical range the localised phase ceases to exist even at arbitrarily strong disorder.
Our results indicate a direct Anderson transition between delocalised and localised phases, with no evidence for an intervening multifractal phase, for both deterministic and random hopping models. 
A scaling analysis based on inverse participation ratios reveals behaviour consistent with a Kosterlitz–Thouless–like transition with two-parameter scaling, in line with Anderson transitions on high-dimensional graphs. We also observe distinct critical behaviour in average and typical correlation functions, reflecting the different scaling properties of generalised inverse participation ratios.
\end{abstract}

\maketitle

\section{Introduction}
The physics of Anderson localisation~\cite{anderson1958absence} in disordered, high-dimensional graphs has been for long, and continues to be, a cornerstone in the context of disordered quantum systems.
This prominence stems in part due to them being fertile ground for studying Anderson transitions in tractable settings~\cite{abou-chacra1973self,economous1972existence,mirlin1991localisation,fyodorov1991localisation}, which in turn offers sharp insights into the nature of the transitions~\cite{tikhonov2016anderson,tikhonov2019critical,tikhonov2019statistics,garciamata2017scaling,garciamata2020two,garciamata2022critical,biroli2022critical,vanoni2024renormalisation} as well as the possibility of realising anomalous multifractal phases~\cite{kravtsov2015random,altshuler2016nonergodic,Tikhonov2016CayleyTree,sonner2017multifractality,kravtsov2018nonergodic}.
This question, more recently, has also attracted renewed attention due to the problem of many-body localisation~\cite{nandkishore2015many,alet2018many,abanin2019colloquium,sierant2025many} being exactly mappable onto an unconventional Anderson localisation problem on the correlated, high-dimensional Fock-space graph~\cite{tikhonov2021anderson,roy2024fock}.

Much of the work on high-dimensional graphs has focussed on two limiting regimes.
The first consists of short-range models, analogous to nearest-neighbour tight-binding Hamiltonians, in which hopping is restricted to pairs of sites connected by edges of the underlying graph~\cite{abou-chacra1973self,mirlin1991localisation,fyodorov1991localisation,monthus2008anderson,monthus2011anderson,biroli2022critical,tikhonov2016anderson,tikhonov2019critical,tikhonov2019statistics,garciamata2017scaling,garciamata2020two,garciamata2022critical,vanoni2024renormalisation,pino2020scaling,rizzo2024localized}.
The second is the case of all-to-all graphs where the hopping amplitude between any two pairs of sites on the graph is statistically the same, such as the Rosenzweig-Porter model and variants thereof~\cite{rosezweig1960repulsion,kravtsov2015random,biroli2021levy,chen2024describing,chen2024quantum,safonova2025spectral}.

However, it has long been appreciated in finite-dimensional systems that spatially decaying but long-ranged hopping, particularly with power-law profiles, can engender a rich array of physical phenomena. These include but are not limited to robust multifractality such as those in power-law banded random matrices~\cite{mirlin2000multifractality,evers2000fluctuations}, cooperative shielding in homogeneous long-ranged hopping models~\cite{celardo2016shielding}, emergent multifractality from algebraic localisation~\cite{deng2018duality,das2025emergent}, and Anderson transitions or absence of localisation altogether due to the interplay of the power-law decay exponent with the spatial dimensionality of the lattice~\cite{logan1985anderson,logan1987dipolar,burin1989localization,levitov1989absence,levitov1990delocalisation,mirlin1996transition,nosov2019robustness,deng2022anisotropy}. 

\begin{figure}
    \centering
    \includegraphics[width=\linewidth]{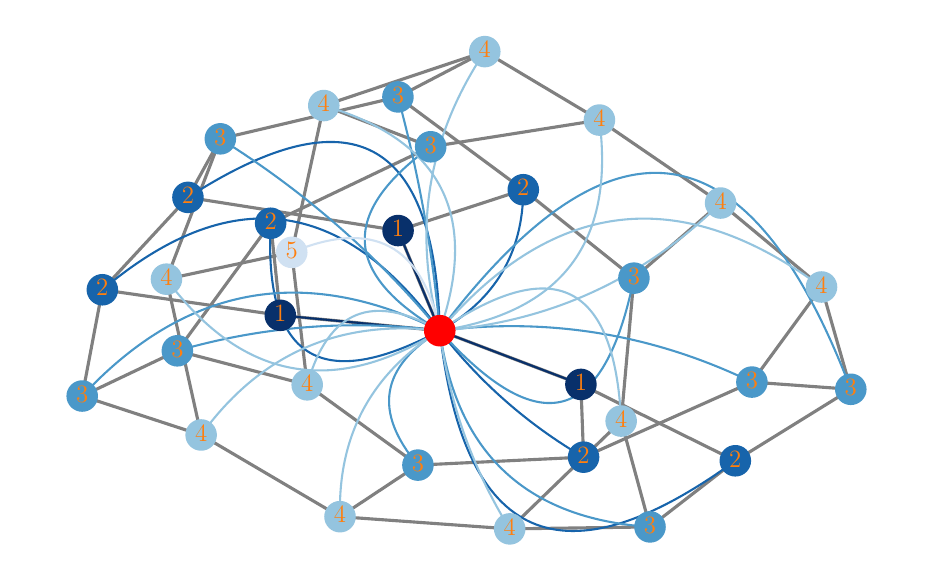}
    \caption{Schematic representation of the ExpRRG model. The underlying graph is the skeleton RRG with connectivity $K=3$. To represent the spatially structured long-ranged hopping, we consider a reference site (marked as red) and the rest of the sites are coloured according to their distance from the reference with lighter colours representing farther distances. The long-ranged hoppings indicated by the additional edges are also coloured according to their strength with darker colours representing larger hopping amplitudes.}
    \label{fig:exprrg-illustration}
\end{figure}

This naturally raises the question of the fate of Anderson localisation on high-dimensional graphs endowed with spatial structure through long-ranged but distance-dependent hopping amplitudes.
In this work, we address this question by introducing and studying a class of models that fill the gap between the extensively studied short-range Anderson model on a random regular graph (RRG) and fully connected models with statistically uniform hopping.
Our construction may be viewed as a power-law generalisation of the Anderson model on an RRG: we embed an instance of an RRG into a complete graph, thereby endowing the graph with a natural notion of distance, and allow the hopping amplitude between any two sites to decay exponentially with their separation, characterised by a length scale~$\xi$ (see Fig.~\ref{fig:exprrg-illustration}).
This framework provides a controlled setting in which the range of hopping can be tuned continuously, while retaining the underlying graph connectivity, and thereby enables a systematic exploration of how spatially decaying long-range hopping modifies localisation, ergodicity, and multifractality in high-dimensional graphs.
We consider both deterministic hoppings, for which the amplitude depends only on the inter-site distance, and random hoppings, where the amplitudes are random but drawn from a distribution whose envelope decays with distance.

We find a nontrivial interplay between the length scale~$\xi$ and the onsite disorder strength (measured relative to the typical hopping amplitude), resulting in a rich localisation phase diagram for this family of models.
Specifically, the critical disorder strength for the localisation transition increases with increasing~$\xi$.
Remarkably, beyond a critical value of~$\xi$, the localised phase ceases to exist altogether, even at arbitrarily strong disorder.
Our results further indicate that the transition between delocalised and localised phases is a direct Anderson transition, with no evidence for an intervening multifractal phase.
\new{
In addition, numerical scaling analysis suggests that the Anderson transition is governed by a Kosterlitz-Thouless-like universality reminiscient of Anderson transitions on high-dimensional graphs~\cite{garciamata2017scaling,garciamata2020two,garciamata2022critical} and the MBL problem on the Fock-space graph~\cite{mace2019multifractal,roy2021fockspace,sutradhar2022scaling,ghosh2025scaling,roy2025spectral}

\subsection{Overview}
We start with a brief overview of the paper and its main results.
We define concretely the family of our models, which we christen, ExpRRG and ExpRRG-RH in Sec.~\ref{sec:models}.
The nomenclature for the former is motivated by the fact that in the ExpRRG model, the hopping amplitude between two sites decays exponentially with the distance between the two sites, and the distance is defined as the shortest path length between the two sites on an underlying RRG backbone on the graph. 
The ExpRRG-RH variant of the model follows a similar construction but the hoppings are random with a distribution with mean zero and standard deviation which decays exponentially with the distance.

Following a brief review of the  diagnostics employed to chart out the phase diagram in Sec.~\ref{sec:diagnostics},
we present the results for the ExpRRG model in Sec.~\ref{sec:exprrg}.
We establish the phase diagram using a combination of numerical exact diagonalisation and analytical renormalised perturbation theory.
In the numerical analysis, we compute diagnostics including level-spacing ratios, inverse participation ratios (IPRs) of eigenstates, and spatial correlation functions in eigenstates and associated correlation lengths.
Complementarily, within the analytical approach we compute, in a self-consistent manner, the imaginary part of the local self-energy using a renormalised perturbative expansion based on the locator expansion~\cite{Economoubook}.
We find reasonably good agreement between the critical lines obtained from these two approaches (see, e.g., Fig.~\ref{fig:pd-exprrg}).
We also show numerically, in Sec.~\ref{sec:exprrg-rh}, that the qualitative features of the phase diagram remain the same for the ExpRRG-RH model.

They two key features of the phase diagram are: (i) With increasing $\xi$ the critical disorder strength increases but there exists a critical values of $\xi$ beyond which even an arbitrarily large disorder strength fails to localise the system in the thermodynamic limit. This is essentially due to the fact that for large enough $\xi$, the exponential decay of the hopping is not fast enough to compete with the exponential growth of the number of sites at a fixed distance. (ii) The Anderson transition is a direct transition between delocalised and localised phases with no intervening multifractal phase.
This observation is particularly notable for the random-hopping model: while randomly weighted Erd\H{o}s-R\`enyi graphs are known to host robust multifractality due to the presence of weak links~\cite{CugliandoloPRB2024}, the fixed underlying topology in the models studied here suppresses the formation of such weak links, leading instead to a sharp transition between delocalised and localised phases controlled by the disorder strength and~$\xi$.

We further analyse the critical behaviour numerically using a scaling theory based on inverse participation ratios (IPRs) in Sec.~\ref{sec:critical}.
Our results are consistent with a Kosterlitz–Thouless–like scenario for the transition, characterised by two-parameter scaling, in line with the generic nature of Anderson transitions on high-dimensional graphs~\cite{garciamata2022critical}.
This behaviour manifests through the emergence of a characteristic value~$q_\ast$ separating two regimes of the generalised $q$th-order IPRs.
For $q>q_\ast$, the IPR exhibits volumic scaling on the ergodic side of the transition and linear scaling in the localised phase, whereas for $q<q_\ast$ it displays linear scaling on both sides of the transition.
The former regime is dominated by rare, high-amplitude peaks in the wavefunction and suppresses contributions from low-intensity regions, and is therefore related to disorder-averaged correlation functions.
In contrast, the latter probes the typical, low-amplitude structure of the wavefunction and is thus connected to typical correlation functions.
The different scaling laws for large and small $q$ are concomitant with the distinct critical behaviour of average and typical correlation functions
}

\subsection{Organisation of the paper}

The remainder of the paper is organised as follows.
In Sec.~\ref{sec:models}, we define the models in detail and discuss how they constitute power-law hopping generalisations of the Anderson model in random regular graphs.
In Sec.~\ref{sec:diagnostics}, we briefly recapitulate the diagnostics used to characterise the localisation phase diagram and the nature of the Anderson transition.
Section~\ref{sec:exprrg} presents the analysis and results for the model with deterministic hoppings, beginning with numerical exact diagonalisation studies and followed by an analytical self-consistent theory for the phase diagram.
This is followed by results for the model with random hoppings in Sec.~\ref{sec:exprrg-rh}.
Finally, we conclude with a summary and a discussion of open questions in Sec.~\ref{sec:summary}.

\section{Random graphs with spatial structure \label{sec:models}}
We will endow the random graphs with a spatial structure, specifically, in the hopping amplitudes thereon. 
To do so, we embed an RRG of connectivity $K$ into a random graph which naturally implements a notion of distance on the graph.
Between two sites $x$ and $y$, the distance $r_{xy}$ is then defined as the length of the shortest path between the sites following the RRG skeleton. 
With the distances defined in this way, we will consider all-to-all hopping on the graph where the hopping amplitude $t_{xy}$ between sites $x$ and $y$, which could be random, typically has a decaying envelope with the distance $r_{xy}$ between the sites.

Formally, the tight-binding Hamiltonian on the RRG can be written as 
\eq{
H = \underbrace{\sum_{x,y}t_{xy}[\ket{x}\bra{y}+{\rm h.c.}]}_{H_{\rm hop}} + \underbrace{\sum_x\epsilon_x\ket{x}\bra{x}}_{H_{\rm dis}}\,,
\label{eq:ham-gen}
}
where the hopping amplitude 
$t_{xy}\sim P_{t,r}$\,,
is a random number drawn from a distribution which depends explicitly on the distance $r_{xy}$. 
The onsite disorder potential $\epsilon_x\sim {\cal N}(0,W)$ is normally distributed with $W$ the disorder strength.

In this work, we will consider models where the hopping profile decays exponentially with $r$.
This is due to the fact that on the RRG skeleton, the number of sites at distance $r$ from any given site increases exponentially with $r$. 
Hence, for the hopping part of the Hamiltonian, $H_{\rm hop}$, to compete with the disordered potential part of the Hamiltonian, $H_{\rm dis}$, the hopping profile must decay at least exponentially fast in $r$; anything slower leads to the $H_{\rm hop}$ completely overwhelming $H_{\rm dis}$ leading to trivial delocalisation. 

Interestingly, our construction constitutes the generalisation of power-law banded models in finite dimensions to the case of infinite dimensions. 
To see this, note that in finite dimensional lattices (with dimension $d$), the number of sites at distance $r$ scales as $N_r\sim r^{d-1}$. 
If the hopping between sites at distance $r$ then also scales as a power law, $t_r\sim r^{-\alpha}$ which in turn implies that $t_r \sim N_r^{-\alpha/(d-1)}$.
The upshot of this is that the hopping amplitude at distance $r$ scales with the number of sites at distance $r$ as a power law. 
On the other hand, in infinite-dimensional graphs such as RRGs, $N_r$ scales exponentially with $r$ as $N_r\sim e^{G r}$ where $G$ is a constant fixed by the topology of the graph. In this case, if the hopping decays exponentially, $t_r \sim e^{-r/\xi}$, it effectively means $t_r\sim N_r^{-1/G\xi}$ such that $t_r$ again scales with $N_r$ as a power law.
This demonstrates the aforementioned point that our models constitute the power-law banded versions of high-dimensional graphs. 

In the following, we briefly but concretely define the two models that we study in this work.

\subsection{ExpRRG}
The ExpRRG model is defined via
\new{
\eq{
P_{t,r}(t_{xy}) =   \delta(t_{xy}-t e^{-(r_{xy}-1)/\xi})\,.
}}
In this case, the hopping amplitude decays exponentially with the distance between the sites but there is no randomness in them -- the hopping amplitude between any two pair of nodes at distance $r$ from each other is given by $te^{-(r-1)/\xi}$. The exponential decay of the hopping amplitude with the distance motivates the nomenclature ExpRRG.
The model therefore has two parameters, $\xi$ and $W/t$, the localisation phase diagram over which will be studied in Sec.~\ref{sec:exprrg}. 
It is expected that the qualitative features of the phase diagram remain the same if $P_{t,r}$ is not a Dirac-delta function but a sufficiently fast decaying function away from the mean. 
Also note that, in the limit of $\xi\to0$, one recovers the conventionally studied RRG (with nearest neighbour hoppings). 

\subsection{ExpRRG-RH}
The ExpRRG-RH model is a close cousin of the ExpRRG model where the hoppings are random; this motivates the nomenclature ExpRRG-RH (random hoppings).   
Specifically, the hoppings are drawn from a Gaussian distribution
\eq{
P_{t,r}(t_{xy}) =  \frac{1}{\sqrt{2\pi\theta^2_{r_{xy}}}}\exp\left[-\frac{t_{xy}^2}{2\theta^2_{r_{xy}}}\right]\,,
\label{eq:exprrg-rh-dist}
}
where $\theta_r$ decays exponentially with $r$ as 
\eq{
\theta_r = t e^{-(r-1)/\xi}\,.
\label{eq:theta-r}
}
While the average magnitude of the hopping amplitude between two sites separated by a distance $r$ is the same in this case as the ExpRRG model, a key difference is that the distribution of the hoppings has a finite weight (in fact, maximum weight) at vanishingly small values. 
This opens up the possibility of having weak links in the graph, which in turn can lead to multifractality, much in the same way as in randomly-weighted Erd\H{o}s-R\'enyi graphs~\cite{CugliandoloPRB2024}, in addition to the usual localised and delocalised phases.
This raises a question about the fate of multifractality in the presence of the all-to-all topology of the graph with the spatial structure -- we will answer this in detail in Sec.~\ref{sec:exprrg-rh}.

\section{Diagnostics for the phase diagram \label{sec:diagnostics}}
In this section, we outline the main diagnostics that we employ to the chart out the localisation phase diagram and characterise the phases in the models. 
The three central quantities of interest will be the level statistics, specifically the mean level spacing ratio, the generalised inverse participation ratios (IPRs), and spatial correlations of the eigenstates on the graph which are also related to the (non-local) Green's functions.

\subsection{Level statistics}

The universal properties of the energy level statistics has long been used as a powerful diagnostic for distinguishing delocalised, ergodic phases of disordered Hamiltonians from non-ergodic, localised ones~\cite{berry1977level,bohigas1984characterisation}. 
The former is expected to be captured by random matrix theory (RMT) of the symmetry class whereas the latter is expected to mirror the case of uncorrelated energy levels. 
This is rooted in the fact that in an ergodic phase, the energy levels repel each other whereas in a localised phase, since the eigenstates with nearby energies are statistically localised in different regions of the graph, they behave as uncorrelated. 
A useful quantitative diagnostic of this is the mean level-spacing ratio defined as follows. 
Consider the eigenvalues, $\{E_n\}$, of the Hamiltonian to be ordered such that $E_{n+1}\ge E_{n}$ and define $s_n = E_{n+1}-E_n$. The level spacing ratio is defined as 
\eq{
r_n = \frac{\min[s_n,s_{n-1}]}{\max[s_n,s_{n-1}]}\,.
\label{eq:lsr}
}
In the delocalised phase the distribution of $r_n$ follows the universal Wigner-Dyson surmise corresponding to the appropriate symmetry class. The Hamiltonians considered here, if ergodic, fall in the Gaussian orthogonal symmetry class, for which the mean level-spacing ratio, $\braket{r} \approx 0.53$ whereas in the localised phase, the Poisson distribution of $r_n$ yields $\braket{r} = 2\ln2-2\approx 0.386$.
In numerical calculations, it is customary to determine the transition between localised and delocalised phases via a crossing (albeit with some drift) of the data for $\braket{r}$ for different system sizes when plotted as a function of the tuning parameter driving the transition. 

\subsection{Inverse participation ratios}
The statistics of eigenstate amplitudes offer complementary, and arguably richer, information about the nature of the eigenstates from the point of view of their localisation properties.
The moments of the eigenstate amplitudes on the graph are quantified via the generalised inverse participation ratios.
The generalised $q^{\rm th}$-IPR for a state $\ket{\psi}$ on a graph with $N$ nodes is defined as~\cite{evers2008anderson}
\eq{
{\cal I}_q(\ket{\psi}) = \sum_{x=1}^N |\braket{x|\psi}|^{2q}\,.
\label{eq:ipr-gen-q}
}
Of particular importance is the scaling of ${\cal I}_q$ with the system size $N$,
\eq{
{\cal I}_q\sim N^{-\tau_q}\,,
}
where $\tau_q$ is related to the multifractal dimension $D_q$ as $D_q = \tau_q/(q-1)$. 
Delocalised states are indicated by $D_q=1$ for all $q$ whereas localised states are signified by $D_q=0$ for all $q>1$. 
In fact, as we will see shortly, in the localised phase $D_q=0$ for all $q>q_\ast$ where the disorder-strength dependent $q_\ast<1$ encodes some features of the anatomy of the localised eigenstates. 
Multifractality is betrayed by a non-trivial behaviour of $D_q$ with $q$. For instance, physically, $N^{D_2}$ can be interpreted as the number of nodes of the graph on which the wavefunction has support; hence $0<D_2<1$ corresponds to a situation where the wavefunction is supported on an infinite number of nodes in the thermdoynamic limit but on a fraction that vanishes in the thermodynamic limit; this is a characteristic feature of multifractal states. 
For our numerical analysis on finite-sized graphs, we define a flowing (with $N$) multifractal exponent as 
\eq{
\tilde{\tau}_q(N) = \frac{\ln \braket{{\cal I}_q(N/2)} - \ln \braket{{\cal I}_q(N)}}{\ln 2}\,,
\label{eq:flowing-fractal-exponent}
}
where $\braket{\cdots}$ denotes the average over realisations of the RRG and disorder, and  eigenstates in the middle of the spectrum, 
and study the drift of $\tilde{\tau}_q(N)$ with $N$ to understand the results in the thermodynamic limit. 
 
\subsection{Spatial correlations of eigenstates}
While spectral statistics and multifractal analyses together offer unambiguous identification of the phases with regard to their localisation properties, much finer information of the anatomy of the wavefunctions can be obtained from spatial correlations of the eigenstate amplitudes.
In particular, these are important characterisations of the phases themselves and carry insights about the mechanisms underlying the Anderson transitions in high-dimensional graphs. 

In such graphs, the number of nodes at distance $r$ from a given node scales exponentially with $r$. However, the distribution of the eigenstate amplitudes on these set of nodes can be extremely heterogeneous~\cite{garciamata2017scaling,garciamata2020two}.
This can have the effect that on rare set of nodes following a particular branch of the graph, the eigenstate amplitudes are large implying a large correlation length, whereas on typical branches the amplitudes are vanishingly small implying a short correlation length. 
This raises the question therefore that if and how are the Anderson transitions in such models dominated by the rare branches. 

To address this question, we will study, two correlation functions.
The first, defined as the average correlation function, is given by
\eq{
C_{\rm avg}(r,N) = \left\langle\frac{1}{N}\sum_x\frac{1}{N_r(x)}\sum_{y: r_{xy}=r}|\psi_x|^2|\psi_y|^2\right\rangle\,,
\label{eq:Cavg}
}
where $N_{r}(x)$ denotes the number of nodes at distance $r$ from node $x$ and the second sum is over all nodes $y$ which are at distance $r$ from $x$.
Note that the IPR is related to the average correlation function as 
\eq{
C_{\rm avg}(0,N) = N^{-1}\braket{{\cal I}_2}\,.
\label{eq:C0-I2}
}
In addition, $C_{\rm avg}(r,N)$ also satisfies a sum rule. Assuming that the number of nodes, $N_r$ at distance $r$ from a given node is independent of the source node, the normalisation of the wavefunction leads to
\eq{
\sum_{r=0}^{r_{\rm eff}}N_r C_{\rm avg}(r,N) = N^{-1}\,,
\label{eq:Cavg-sumrule}
}
where $r_{\rm eff}$ is the effective diameter of the graph which we will define precisely in the next section.
To define a typical correlation function, we follow the prescription from Ref.~\cite{garciamata2020two,garciamata2022critical} and define it as
\eq{
C_{\rm typ}(r,N) = \exp\left\langle\ln\left(\frac{1}{N}\sum_x|\psi_x|^2|\psi_{x_r}|^2\right)\right\rangle\,,
\label{eq:Ctyp}
}
where $x_r$ is a node chosen randomly over the $N_{r}(x)$ nodes at distance $r$ from $x$.

\subsection{Local Green's functions}
The local Green's function is an important quantity of interest as two related diagnostics of localisation, namely the local density of state (LDoS), and the imaginary part of the local self-energy.
With the Green's function operator defined as $\hat{G}(\omega) = (\omega^+\mathbb{I} - H)^{-1}$, the local Green's function at site $x$ is given by
\eq{
G_{xx}(\omega) = \braket{x|(\omega^+-H)^{-1}|x} = \sum_{E}\frac{|\braket{x|E}|^2}{\omega^+-E}\,,
\label{eq:Gxx}
}
where $\omega^+ = \omega+i\eta$ with $\eta=0^+$ the regulator, and $\{\ket{E}\}$ denotes the set of eigenstates of $H$ with the corresponding eigenvalues being $E$.

The LDoS at site $x$ can be obtained from the local Green's function via\footnote{This is straightforward to see as  ${\rm Im}[G_{xx}(\omega)] = \sum_{E}\frac{-\eta |\braket{x|E}|^2}{(\omega-E)^2 + \eta^2}$ such that $\lim_{\eta\to 0}{\rm Im}[G_{xx}(\omega)] = -\pi\sum_{E}|\braket{x|E}|^2\delta(\omega-E)$ which is nothing but $-\pi D_{x}(\omega)$}
\eq{
D_x(\omega)&=\sum_{E}|\braket{x|E}|^2\delta(\omega-E)\label{eq:Dx-def}\\
&= -\frac{1}{\pi}\lim_{\eta\to 0}{\rm Im}[G_{xx}(\omega)]\,.\label{eq:Dx-G}
}
In a delocalised phase in the thermodynamic limit, the LDoS is continuous function of $\omega$ as eigenstates at all energies overlap any given site. 
On the other hand, in the localised phase, the LDoS is a singular function as on a given site, only those few $O(1)$ eigenstates overlap the site  whose localisation centres are within the localisation length of the given site.
These states have a large contribution to the LDoS whereas the overwhelmingly large number of states have vanishing overlap on the given site leading a vanishing LDoS for most $\omega$.
This qualitative difference between the LDoS in the delocalised and localised phases is made quantitative by studying the probability distribution of $D_x(\omega)$ defined via
\eq{
P_D(\rho,\omega) = \Braket{\frac{1}{N}\sum_{x}\delta(-\pi^{-1}{\rm Im}[G_{xx}(\omega)]-\rho)}\,.
}
In the delocalised phase the distribution is well-behaved and independent of $\eta$, in the thermodynamic limit, such that all moments of the distribution are $O(1)$.
By contrast, in a localised phase, the distribution develops power-law tails which are cut off at scales $\rho\sim 1/\eta$.
The cut-off scale can be simply be understood as being set by the broadening of the $\delta$-functions in the LDoS in Eq.~\eqref{eq:Dx-def} by the regulator $\eta$.
This structure of $P_D$ in the localised phase means that the higher moments of the distribution are dominated by the cut-off $\eta$ and therefore they have a non-trivial scaling with $\eta$.
These differences are captured quite conveniently by the typical value of the LDoS defined as
\eq{
\rho_{\rm typ}(\omega) = \exp\left[\int_0^\infty d\rho~(\ln \rho)P_D(\rho,\omega)\right]\,,
}
which constitutes an order parameter for the localisation transition. 
In a delocalised phase $\rho_{\rm typ}\sim O(1)$ whereas in a localised phase $\rho_{\rm typ}\sim \eta$.

{\let\MakeTextUppercase\relax
 \section{ExpRRG Model} \label{sec:exprrg}
}

The ExpRRG model, defined concretely in Sec.~\ref{sec:models}, has effectively three parameters, the connectivity of the underlying RRG denoted by $K$, the onsite disorder strength $W$, and the lenghtscale $\xi$ which controls the decay of the hopping amplitude\footnote{The hopping amplitude $t$ can be set to unity or equivalently, the disorder strength $W$ can be understood to have been taken relative to the hopping strength.}.
For a given $K$, increasing $\xi$ favours delocalisation as it strengthens the hopping between distant sites whereas increasing $W$ is expected to favour localisation thereby leading to competition between the two.
In the following, we will present results for the localisation phase diagram of the model in the parameter space spanned by $\xi$ and $W$.

However, before delving into that, it is important to note an interplay between the connectivity $K$ and the lengthscale $\xi$.
While the hopping amplitude between two sites decays exponentially with the distance between two, the topology of an RRG implies that the number of sites a distance $r$ from a given site, denoted by $N_r$, grows exponentially with $r$ as,
\eq{
N_r = K (K-1)^{r-1}\,;~1\leq r\leq r_{\rm eff}\,,
}
where 
\eq{
r_{\rm eff} = \ln N/\ln(K-1)\,,
}
can be understood as the {\it effective} diameter of the graph. More precisely, $r_{\rm eff}$ is the distance $r$ where $N_r$ is maximised and $N_r$ drops to zero rapidly within a few $O(1)$ distances from $r_{\rm eff}$.
The norm of the kinetic part of the Hamiltonian can therefore be computed as 
\eq{
\braket{||H_{\rm kin}||_2} &= Nt^2\sum_{r=1}^{r_{\rm eff}}N_r e^{-2(r-1)/\xi}\,,\nonumber\\
&=Kt^2N\frac{1-(K-1)^{r_{\rm eff}}e^{-2r_{\rm eff}/\xi}}{1-(K-1)e^{-2/\xi}}\,.
}
The above expression shows that for $\xi > 2/\ln(K-1)$, the norm $\braket{||H_{\rm kin}||_2}\sim N^2$ and the kinetic part of the Hamiltonian ceases to be extensive and becomes super-extensive. In a such scenario, it completely dominates the disordered potential part of the Hamiltonian whose norm,
\eq{
\braket{||H_{\rm pot}||_2} = W^2N
}
scales necessarily as $N$. 
This naturally suggests that for $\xi > 2/\ln(K-1)$, a localised phase cannot exist and in the rest of what follows we will restrict ourselves to $\xi < 2/\ln(K-1)$. 
In this case, the kinetic term has a norm in the limit of $N\gg 1$, given by
\eq{
\braket{||H_{\rm kin}||_2} \overset{N\gg1}{\approx} Kt^2N\frac{1}{1-(K-1)e^{-2/\xi}}\,.
}
More importantly, the norms of the both the kinetic and potential energy parts of the Hamiltonian are now extensive. 
As such, one can estimate a loose bound for the localised phase as being given by $\braket{||H_{\rm pot}||_2}>\braket{||H_{\rm kin}||_2}$, which is the condition that 
\eq{
W>\frac{Kt}{\sqrt{1-(K-1)e^{-2/\xi}}}\,.
}
However, as we will show next using a combination numerical and analytical results, the actual critical disorder strength lies much above this bounding value for a given value of $K$ and $\xi$.

\subsection{Numerical results}
\begin{figure}[!b]
    \centering
    \includegraphics[width=\linewidth]{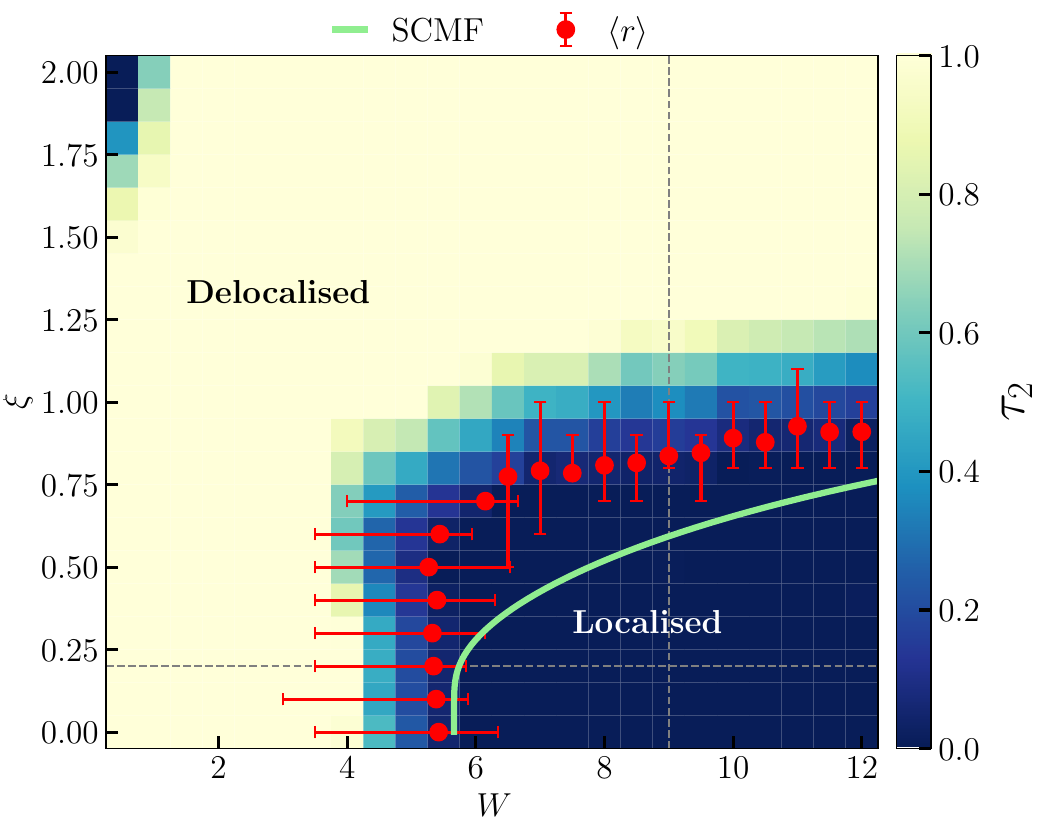}
    \caption{Phase diagram of the ExpRRG model obtained numerically from ED in the $W$-$\xi$ plane. The heatmap shows the IPR exponent, $\tau_2$ with darker colours denoting $\tau_2\approx 0$ in the localised phase and lighter colours corresponding to $\tau_2\approx 1$ in the delocalised phase. The red circles denote the critical points obtained from the analysis of the level statistics and the green line shows the locus of critical points obtained analytically from a self-consistent mean-field theory in Sec.~\ref{sec:selfconsisten}.}
    \label{fig:pd-exprrg}
\end{figure}

In this section, we discuss the numerical results obtained using exact diagonalisation (ED) for $K=3$. Specifically, we focus on $N/32$ eigenstates in the middle of the spectrum, and our results are averaged over these eigenstates as well as approximately 10000-500 realisations of RRGs and disorder for the smallest and largest sizes. 
The phase diagram of the model, obtained from the various diagnostics computed using ED is shown in Fig.~\ref{fig:pd-exprrg}.
In the following, we describe the behaviour of the diagnostics described in Sec.~\ref{sec:diagnostics}, which we employ the obtain the phase diagram.

\subsubsection{Phase diagram}

\begin{figure}[!t]
    \centering
    \includegraphics[width=\linewidth]{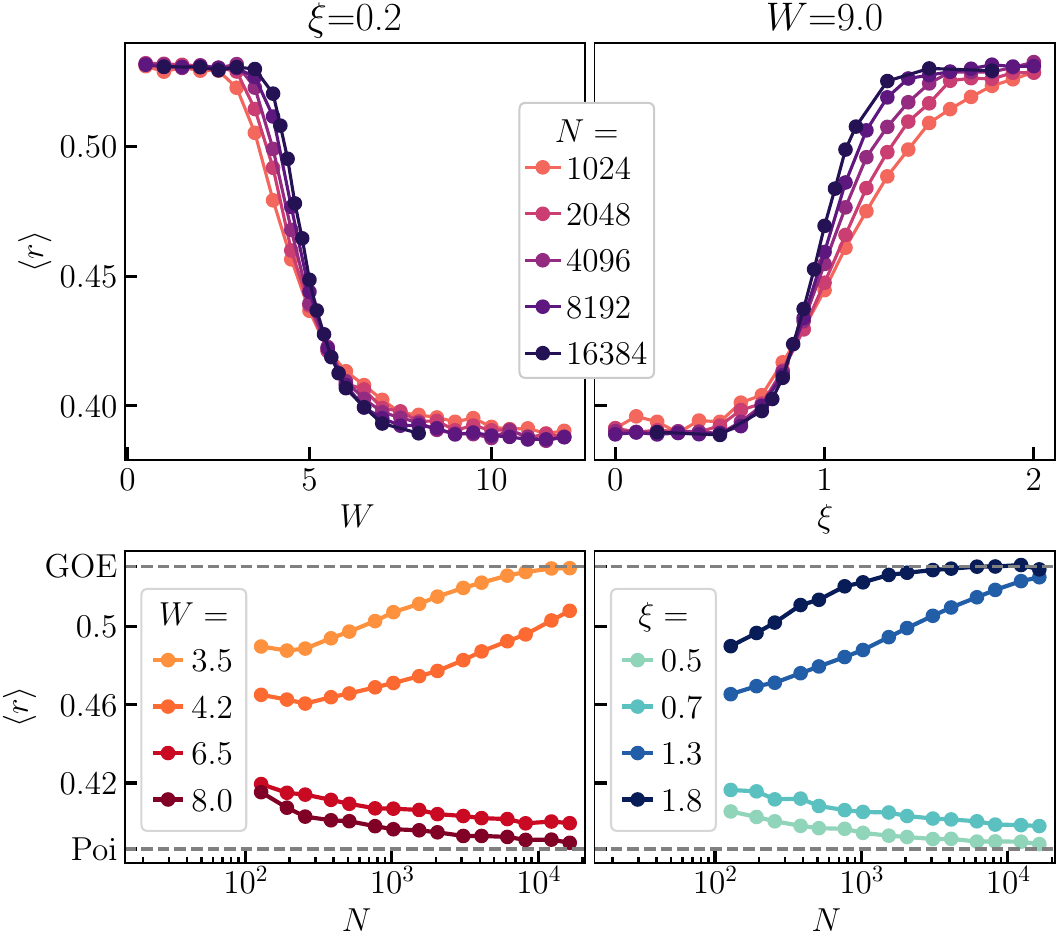}
    \caption{Level statistics for the ExpRRG model. The top panels shows the mean level spacing ratio, $\braket{r}$, as a function of $W$ for a fixed $\xi$ (left) and as a function of $\xi$ for a fixed $W$ (right), for different system sizes. These correspond to the horizontal and vertical cuts shown by the dashed lines in Fig.~\ref{fig:pd-exprrg}. The crossing in the data for different $N$ indicates the Anderson transition, and crossings such as the ones seen here were used to estimate the phase boundary in Fig.~\ref{fig:pd-exprrg}. The lower panels show the drift of the data with $N$ at parameter values representative of the delocalised and localised phases. For the former, the data approaches the GOE value of $\approx 0.53$ with increasing $N$ whereas for the latter it approaches the Poisson value of $0.386$.}
    \label{fig:lstats-exprrg}
\end{figure}

We start with the mean level-spacing ratio, defined in Eq.~\ref{eq:lsr}, representative results for which are shown in Fig.~\ref{fig:lstats-exprrg}.
The upper left panel shows $\braket{r}$ as a function of disorder strength $W$ for a fixed value of $\xi$. 
The data clearly goes from  the GOE values of $\approx 0.53$ at weak disorder to the Poisson value of $0.386$ at strong disorder, with the crossover getting sharper with increasing system size $N$. The transition is indicated by a crossing of the data for different $N$. The data corresponds to the horizontal slice (grey dashed line) in the phase diagram in Fig.~\ref{fig:pd-exprrg}. The lower left panel shows the dependence of $\braket{r}$ with $N$ for two representative point each on the delocalised and localised sides of the transition. For the former (latter), with the increasing $N$, the data approaches the asymptotic GOE (Possion) value.

Analogously, in the right panels in Fig.~\ref{fig:lstats-exprrg}, we show the data for $\braket{r}$ as a function of $\xi$ at a fixed $W$ corresponding to the vertical cut (grey dashed line) in the phase diagram. The phenomenology is identical as before with the data transitioning from the localised Poisson value at small values of $\xi$ to delocalised GOE value at larger values of $\xi$ with the transition at a critical $\xi$ again indicated by the crossing of the data for different $N$.

By tracking the crossing points of the data for $\braket{r}$ along various vertical and horizontal cuts in the $W$-$\xi$ plane, we obtain a phase boundary between the localised and delocalised phases in the model.
This is shown by the red circles in Fig.~\ref{fig:pd-exprrg}.
We take a conservative approach towards the estimating the errorbars by considering the parameter values near the critical point where $\braket{r}$ grows or decays monotonically with $N$ as the bounds for the delocalised and localised phases respectively.

\begin{figure}[!t]
    \centering
    \includegraphics[width=\linewidth]{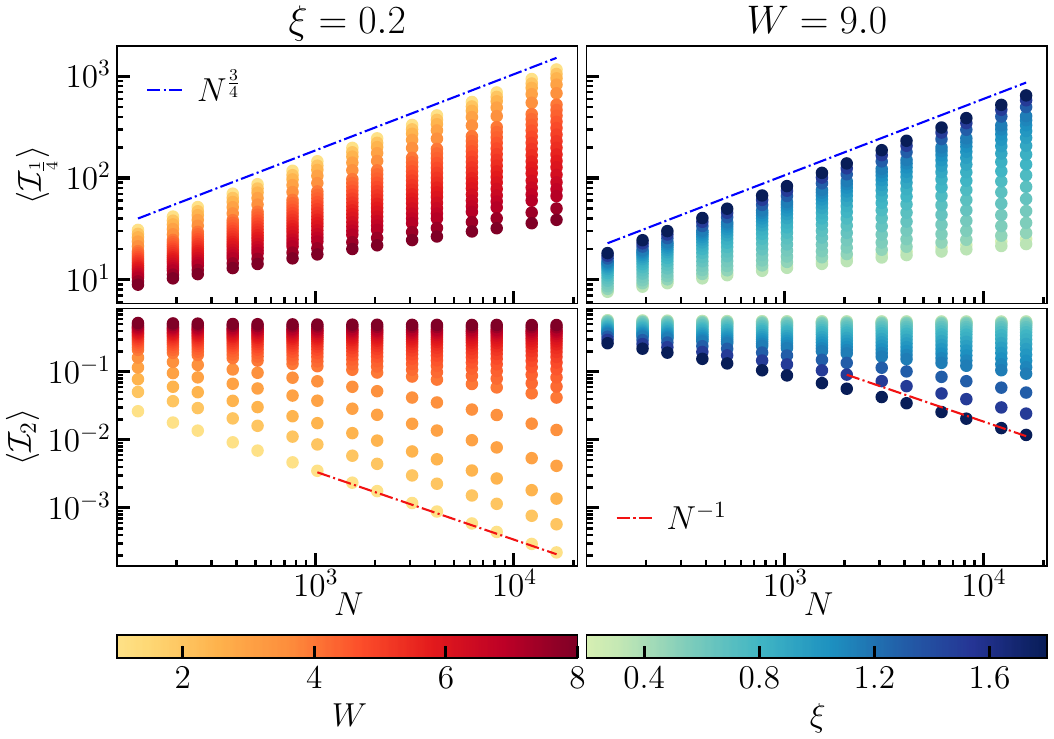}
    \caption{The generalised IPRs, $\braket{I_q}$ for the ExpRRG model, for $q=1/4$ (top) and $q=2$ (bottom). The left panel shows the data as a function of $N$ for different $W$ (different colours) for a fixed $\xi$ whereas in the right panel, different colours denote different $\xi$ values at a fixed $W$. }
    \label{fig:ipr-vs-N}
\end{figure}

We next discuss the statistical properties of the eigenstate wavefunctions on the RRG, as encoded in the IPRs defined in Eq.~\ref{eq:ipr-gen-q}.
While it is quite common to study the case of $q=2$, here we will focus on a range of $q$ focussing on both $q>1$ as well as $q<1$ as the two contain complementary information about the eigenstates. 
Larger values of $q$ amplify the peaks in the wavefunction and therefore probe the number of fraction of sites where it has an overwhelming support.
On the other hand, smaller values of $q$ suppress the peaks and amplify the smaller wavefunction amplitudes, thereby probing the fluctuations in the wavefunctions at smaller scales. In fact, for Anderson localised states, while $\tau_q=0$ necessarily for $q>1$, one can have non-trivial behaviour of $\tau_q$ for $q<1$ (especially $q<1/2$); a phenomenon dubbed `strong multifractality'.
As we will discuss later, this is also connected to the average versus typical properties of the wavefunctions in the localised phase.

In Fig.~\ref{fig:ipr-vs-N} we show the data for $\braket{{\cal I}_q}$ for two representative values of $q=1/4$ and $q=2$ as a function of $N$. In the left column, we show data at a fixed $\xi$ for different values of $W$ whereas in the right panel we show the data at a fixed $W$ for different values of $\xi$; these again correspond to the vertical and horizontal cuts in the phase diagram denoted by the grey dashed lines in Fig.~\ref{fig:pd-exprrg}.
For $q=2$, the data approaches the $\sim N^{-1}$ behaviour at small $W$ for a fixed $\xi$ or large $\xi$ at fixed $W$ whereas it approaches the $\sim N^0$ behaviour large $W$ for a fixed $\xi$ or small $\xi$ at fixed $W$ -- this indicates a transition from a delocalised to a localised phase. 
For $q=1/4$ as well, we see the $\sim N^{-(q-1)=3/4}$  behaviour, characteristic of a delocalised phase at small $W$ and large $\xi$.
On the other hand, in the localised phase at large $W$ and small $\xi$, the data shows that ${\cal I}_{1/4}$ continues to have a non-trivial scaling with $N$ indicating the strong multifractality of the localised phase. 
While the signature of a transition from a localised to a delocalised phase is compelling, there are significant finite-size corrections in the data as manifested in the deviation of the data of $\ln{\braket{{\cal I}_q}}$ from a perfect power-law behaviour with $N$. Hence, to extract the phase diagram we fit the data to a form~\cite{luitz2015many,mace2019multifractal}
\eq{
\ln{\braket{{\cal I}_q}} = \tau_q \ln N + b_q \ln\ln N\,,
}
and the results for $\tau_2$ as heatmap in the $W$-$\xi$ plane is shown in Fig.~\ref{fig:pd-exprrg}. The localised and delocalised phases are indicated by a value of $\tau_2\approx 0$ and $\approx 1$ respectively. 
More importantly the phase boundary between the two phases as indicated by the $\tau_2$ is in good agreement with the prediction from the level statistics within error bars.
This gives further credence to the localisation phase diagram.

\begin{figure}
    \centering
    \includegraphics[width=\linewidth]{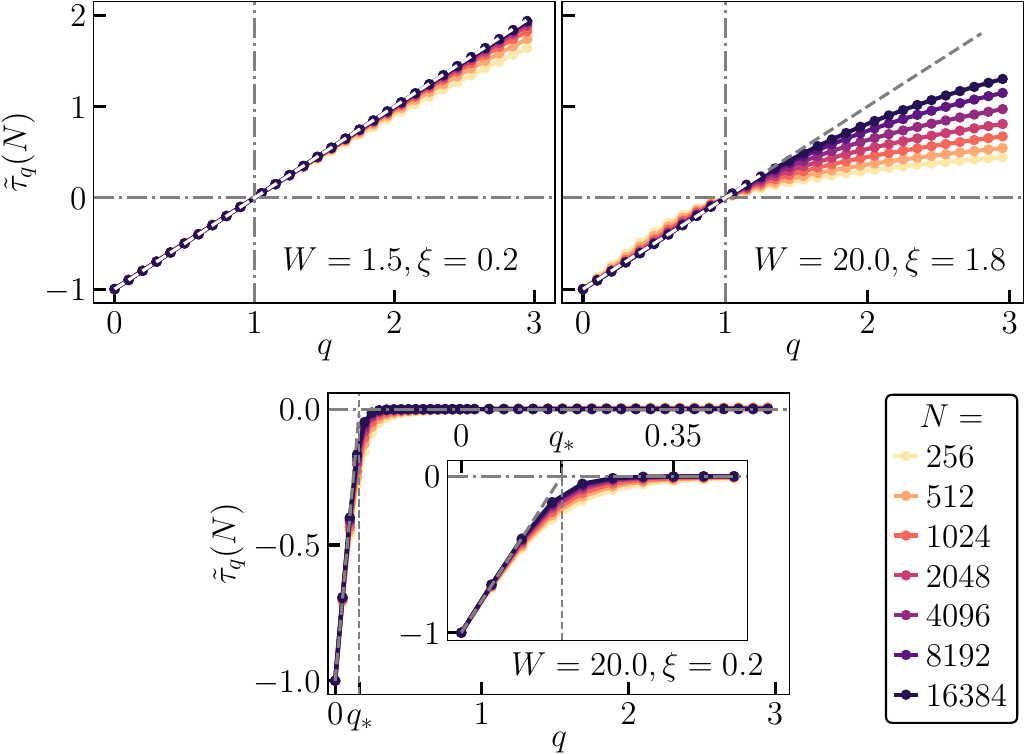}
    \caption{Flowing fractal exponent $\tilde{\tau}_q(N)$ as a function of $q$ (defined in Eq.~\ref{eq:flowing-fractal-exponent}) at  three different points in the parameter space for different $N$. The upper panels correspond to the delocalised phase at weak disorder (left) and strong disorder but large $\xi$ (right). The dashed lines denote the asymptotic $q-1$ behaviour.
    The lower panel corresponds to the localised phase at strong disorder and small $\xi$. The inset shows the same data as the main panel but zoomed in to focus on the low $q$ regime. The red dashed line shows the fit to form in Eq.~\ref{eq:qstar} which is used to extract $q_\ast$.
    }
    \label{fig:flowing-tau-q-at-three-points}
\end{figure}

While the scaling of $\braket{{\cal I}_2}$ and the resulting exponent $\tau_2$ is sufficient to establish the localisation phase diagram, further insights into the finer structures of the phases are obtained by studying $\tau_q$ as a function of $q$.
In Fig.~\ref{fig:flowing-tau-q-at-three-points} we show the flowing multifractal exponent $\tilde{\tau}_q$, defined in Eq.~\ref{eq:flowing-fractal-exponent}, as a function of $q$ for different values of $N$ at three representative points in the phase diagram. 
The upper left panel shows the case of delocalised phase at weak disorder ($W=1.5$) where the data rapidly falls onto the $q-1$ behaviour, characteristic of the delocalised phase.
The delocalised phase at strong disorder ($W=20$) but also at large values of $\xi$ ($\xi=1.8$), has much stronger finite-size effects and the convergence to the $q-1$ behaviour is much slower with $N$, as evident in the uppper right panel. More importantly, we do not find an extended regime in our parameter space where $\tau_q$ converges to a non-trivial function of $q$ different from the delocalised and localised behaviour. This leads us to conclude that the ExpRRG model as defined with $K=3$ does not host a multifractal phase and there is a direct transition between the delocalised and localised phases; we will return to this issue shortly again.

In the localised phase (bottom panel), with increasing $N$, $\tilde{\tau}_q$ converges to a form which has a non-trivial $q$ dependence for small $q$ and is zero otherwise indicative of strong multifractality. More specifically, the form can be expressed as 
\eq{
\lim_{N\to\infty}\tilde{\tau}_q (N) = \begin{cases}
                                            {q}/{q_\ast(W,\xi)}-1\,;&q\leq q_\ast(W,\xi)\\
                                            0\,;&q>q_\ast(W,\xi)
                                        \end{cases}\,,
\label{eq:qstar}}
where $q_\ast(W,\xi)<1/2$ in the localised phase and as a critical point is approached $q_\ast\to 1/2$. The latter is mandated by a fundamental symmetry of the multifractal spectrum and the scale invariance of the critical point~\cite{mirlin1994distribution,mirlin2006exact,bilen2021symmetry}.

\begin{figure}
    \centering
    \includegraphics[width=\linewidth]{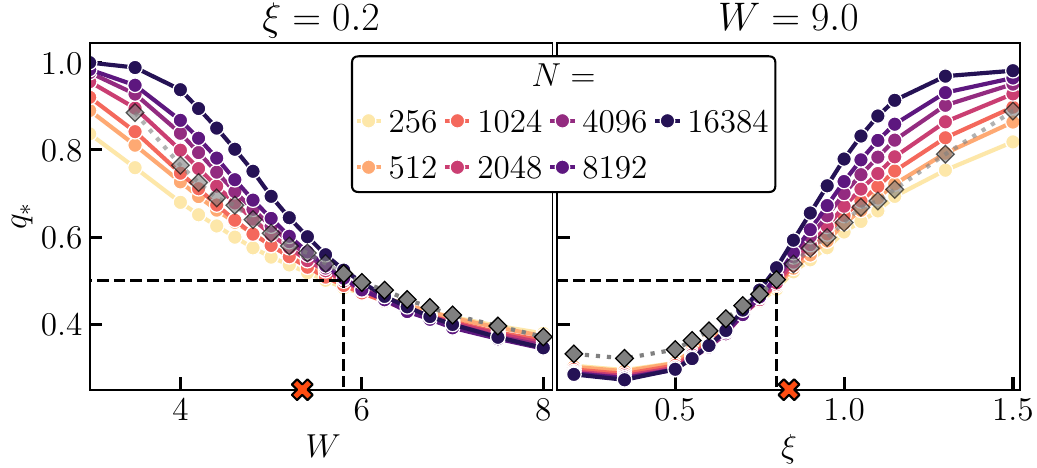}
    \caption{$q_\ast$ as a function of $W$ and $\xi$ along the representative horizontal and vertical cuts in Fig.~ \ref{fig:pd-exprrg} for different $N$. The data is well converged with $N$ in the localised phase and $q_\ast\to 1/2$ as the critical point is approached from the localised side. In the delocalised phase, the data grows with $N$ towards its saturation value of $1$. The grey squares denote $\zeta_{\rm typ}\ln(K-1)$ (obtained from the typical correlation function $C_{\rm typ}$) which shows an excellent agreement with $q_\ast$ corroborating the result in Eq.~\ref{eq:qstar-zeta-typ}. The red cross marks on the horizontal axis denote the critical points obtained from the finite size scaling of $\langle r \rangle$. }
    \label{fig:qstar}
\end{figure}

To extract $q_\ast$ quantitatively, we consider the data for $\tilde{\tau}_q(N)$ in the vicinity of small $q$ and fit it to the form $q/q_\ast -1$. The resulting $q_\ast(W,\xi)$ are shown in Fig.~\ref{fig:qstar} for the representative horizontal and vertical cuts of the phase diagram.
In the delocalised phase, the data rapidly drifts towards $q_\ast=1$ with increasing $N$ consistent with the behaviour that $\tilde{\tau}_q(N)\overset{N\to\infty}{\to} q-1$ for all $q\geq 0$.
On the other hand, in the localised phase, the data for $q_\ast$ is well converged with $N$ to a value which is necessarily $\leq 1/2$ and approaches $1/2$ as the critical point is approached.
In fact, estimating the critical point to be the parameter value where $q_\ast$ approaches $1/2$ is in reasonable agreement with the estimate obtained independent from the analysis of the level statistics. 
The observation that $q_\ast$ approaches $1/2$ as the transition is approached from the localised side and jumps to 1 in the delocalised phase will play an important role in determining the nature of the transition, as we will discuss in Sec.~\ref{sec:critical}.

We briefly return to the point about the absence of a multifractal phase in the ExpRRG model, at least within the realms of our numerical study. 
In terms of $q_\ast$, a genuine multifractal phase would be implied by the convergence of $q_\ast$ with $N$ to value $1/2<q_\ast<1$. However, as shown representatively in Fig.~\ref{fig:qstar}, we find no such regime in our numerical analysis. Wherever $q_\ast$ converges with $N$, it does so to a value $<1/2$ indicative of a localised phase, and otherwise the data grows with $N$ seemingly towards its saturation value of 1.
This provides further suggestive evidence that the ExpRRG model, as defined, with $K=3$ hosts a direct Anderson transition between a delocalised and a localised phase.

\begin{figure}[!t]
    \centering
    \includegraphics[width=\linewidth]{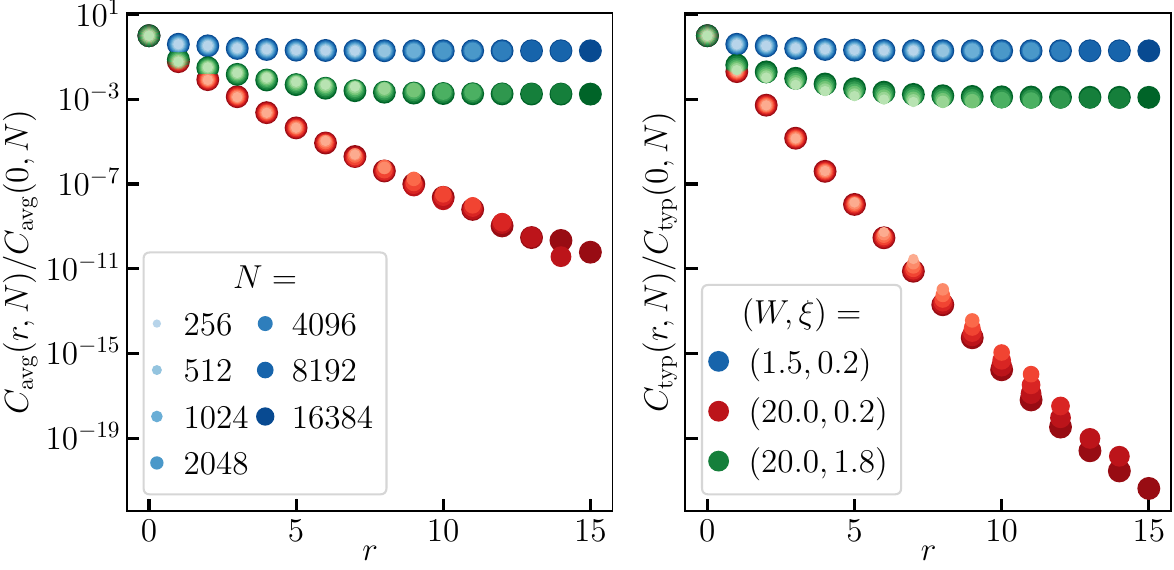}
    \caption{The average and typical correlation functions, defined in Eq.~\ref{eq:Cavg} and Eq.~\ref{eq:Ctyp} as a function of $r$ scaled by their values at $r=0$. The different colours correspond to different points in the parameter space; blue and green correspond to the delocalised phases at weak and strong disorder with the latter at large $\xi$ whereas the red represents the behaviour in a localised phase. For each parameter value, data for different $N$ are denoted by different intensities of colour (and size of data points), and they are converged with $N$.}
    \label{fig:correlation-functions}
\end{figure}

As the final numerical diagnostic, we next discuss the results for the correlation functions defined in Eq.~\ref{eq:Cavg} and Eq.~\ref{eq:Ctyp}.
As discussed earlier, the average and the typical correlation function, $C_{\rm avg}$ and $C_{\rm typ}$ probe two different aspects of the structure of eigenstates.
The results shown in Fig.~\ref{fig:correlation-functions} suggest that they satisfy a form
\eq{
C_{\rm avg/typ}(r,N)  =C_{\rm avg/typ}(r=0,N)g_{\rm avg/typ}(r)\,,
}
where the key point is that the function $g_{\rm avg/typ}$ which encodes the $r$-dependence is independent of $N$.
The data shows that in the delocalised phase $g_{\rm avg/typ}(r)\overset{r\gg 1}{\to} c$ where $c$ is constant with $r$ but depends obviously on $W$ and $\xi$.
On the other hand, in the localised phase, $g_{\rm avg/typ}(r)$ seemingly decays exponentially with $r$
\eq{
g_{\rm avg/typ}(r) = \exp[-r/\zeta_{\rm avg/typ}]\,.
\label{eq:g-def}
}
However, the data also makes it clear that $\zeta_{\rm avg}$ is significantly different from $\zeta_{\rm typ}$.
This is suggestive of the fact that, in the localised phase, there exist rare branches of the graph on which the eigenstates primarily live, $C_{\rm avg}$ and $C_{\rm typ}$ would behave very differently. 
From a given site $x$, there exist $N_r\sim K^r$ sites, say $\{y\}$, at distance $r$. The heterogeneity of the wavefunction would imply that on some of the $y$-sites, the correlation function, $|\psi_x|^2|\psi_y|^2$ is anomalously large in the background of much smaller correlations on the majority of the $y$-sites.
This in turn implies that the typical correlation will be much smaller than the average ones as the latter will be dominated by the rare sites with anomalously large correlations, as evidenced by the data in Fig.~\ref{fig:correlation-functions}
As we will discuss later, the distinction between  $\zeta_{\rm avg}$ and $\zeta_{\rm typ}$ becomes crucial near the Anderson transition and shed important insights onto the nature of the transition.

The correlation lengths, $\zeta_{\rm avg/typ}$, as defined above are intimately connected to the IPRs as follows. 
Consider first the case of $C_{\rm avg}(r,N)$.
Using the Eq.~\ref{eq:C0-I2} and the exponential form in Eq.~\ref{eq:g-def} in the sum rule satisfied by $C_{\rm avg}$ in Eq.~\ref{eq:Cavg-sumrule} we have
\eq{
\sum_{r=0}^{r_{\rm eff}}N_r \exp[-r/\zeta_{\rm avg}] = \braket{{\cal I}_2}^{-1}\,.
}
The sum in the equation can be evaluated exactly to give an expression for $\braket{{\cal I}_2}$ as 
\eq{
\langle&{\cal I}_2\rangle=\left(1+K\frac{N^{1-1/\zeta_{\rm avg}\ln(K-1)}-1}{K-1-e^{1/\zeta_{\rm avg}}}\right)^{-1}\\
&\overset{N\gg 1}{\approx}\begin{cases}
\left(1-\frac{K}{K-1-e^{1/\zeta_{\rm avg}}}\right)^{-1}\,;&\zeta_{\rm avg}<\zeta_{\rm avg}^{c}\\
\frac{K^{-1}N^{-(1-1/\zeta_{\rm avg}\ln(K-1))}}{(K-1-e^{1/\zeta_{\rm avg}})^{-1}}\,;&\zeta_{\rm avg}>\zeta_{\rm avg}^{c}\\
\frac{K-2}{K}N^{-1}\,;&\zeta_{\rm avg}\to\infty
\end{cases}\,,
\label{eq:I2-corr}
}
where $\zeta_{\rm avg}^{c} = 1/\ln(K-1)$.

The equation above provides an explicit relation between the $q=2$ IPR and the correlation length $\zeta_{\rm avg}$.
The key point to notice is that the localised phase exists, signified by an $O(1)$ value of $\braket{{\cal I}_2}$, only for $\zeta_{\rm avg}<\zeta_{\rm avg}^{c}$.
On the other hand, the delocalised phase with $\braket{{\cal I}_2}\sim N^{-1}$ occurs only for for a divergent $\zeta_{\rm avg}$.
For any finite value of $\zeta_{\rm avg}>\zeta_{\rm avg}^c$, the $q=2$ IPR, $\braket{{\cal I}_2}$, exhibits multifractal scaling with $\tau_2 = 1-1/\zeta_{\rm avg}\ln(K-1)$.
However, the ExpRRG model has a direct transition between localised and delocalised phases, without any intervening multifractal phase.
This leads to a key result that at the Anderson transition, $\zeta_{\rm avg}$ does not diverge and instead approaches a finite value of $\zeta_{\rm avg}^{c} = 1/\ln(K-1)$.

As far as the $C_{\rm typ}$ is concerned, it is natural to expect that it will be related to the small-$q$ IPRs.
This is simply because the $C_{\rm typ}$ probes the correlations on typical branches where the wavefunctions are very small. In the same way, the small-$q$ IPRs probe the fluctuations of the wavefunction amplitudes at such small values. In fact, this essential idea has been used to relate $C_{\rm typ}$ and $\zeta_{\rm typ}$ with $\braket{{\cal I}_{q\ll 1}}$~\cite{garciamata2020two,garciamata2022critical}, and here we discuss the results along very similar lines. 
The picture for a localised eigenstate is that there exists a root site $x_{\rm root}$ where the eigenstate is localised and away from it, the amplitude decays typically exponentially with the distance from the root. 
The IPR, valid for small $q$ in such situation is given by
\eq{
{\cal I}_q = \frac{\sum_{r=0}^{r_{\rm eff}}N_r e^{-qr/\zeta_{\rm typ}}}{[\sum_{r=0}^{r_{\rm eff}}N_r e^{-r/\zeta_{\rm typ}}]^q}\overset{\substack{N\gg 1\\q\ll 1}}{\sim} N^{-\left[\frac{q}{\zeta_{\rm typ}\ln(K-1)}-1\right]}\,.
}
At the same time, we also know that in this regime $\tau_q = q/q_\ast -1$, which leads to the identification that 
\eq{
q_\ast(W,\xi) = \zeta_{\rm typ}(W,\xi)\times \ln(K-1)\,.
\label{eq:qstar-zeta-typ}
}
This result is corroborated numerically in Fig.~\ref{fig:qstar} where indeed $\zeta_{\rm typ}\ln(K-1)$ is an excellent agreement with the $q_\ast$ in the localised phase. 
An important fallout of the result is that since $q_\ast\to1/2$ as the critical point is approached from the localised side, $\zeta_{\rm typ}$ also approaches a finite value at criticality given by $\zeta_{\rm typ}^c = 1/2\ln(K-1)$. Although neither $\zeta_{\rm avg}$ or $\zeta_{\rm typ}$ diverge at the critical point and approach finite values $\zeta_{\rm avg/typ}^c$, we will show shortly that the critical exponents obtained from a scaling theory control this approach.

\subsubsection{Critical properties \label{sec:critical}}

\begin{figure}
\includegraphics[width=\linewidth]{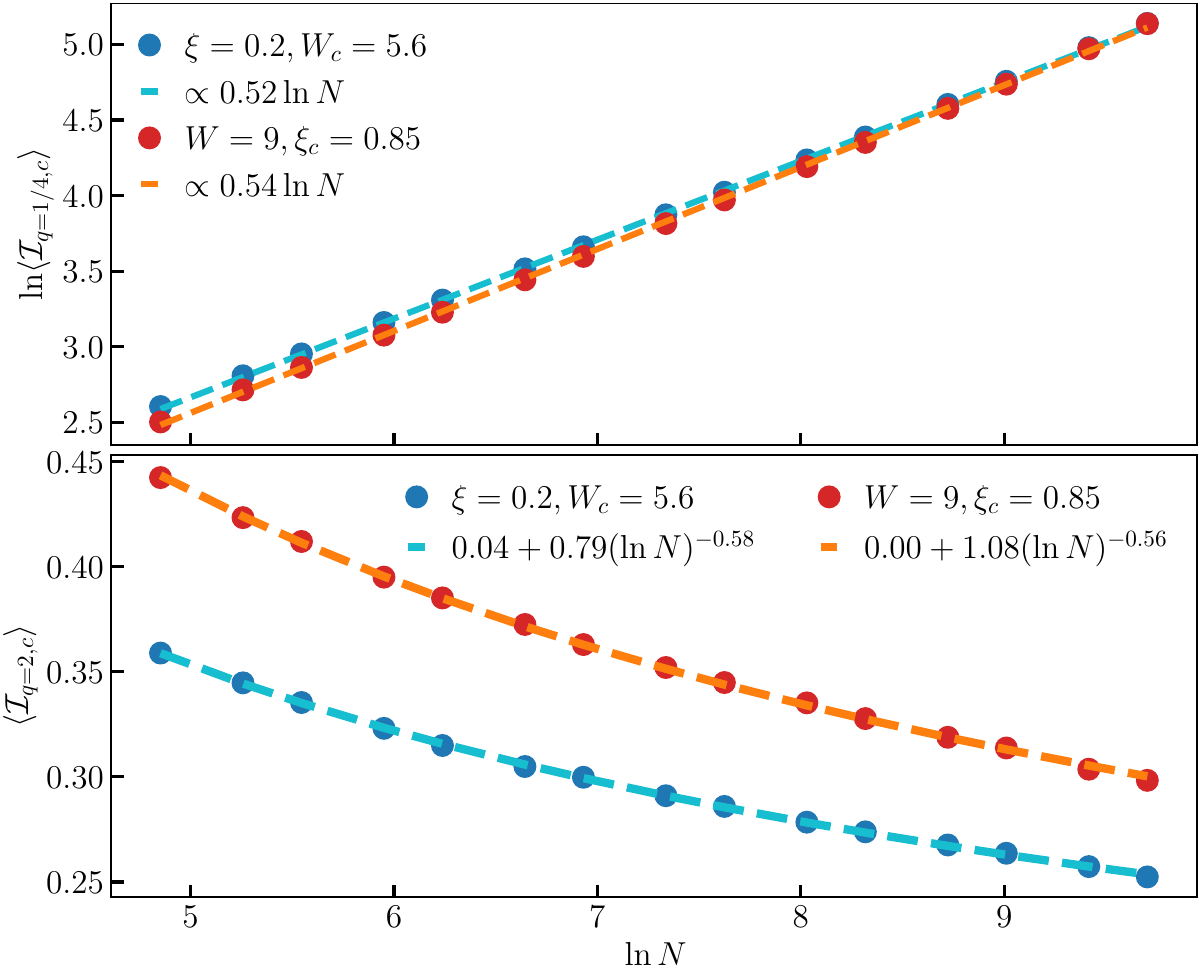}
\caption{The scaling of the $q^{\rm th}$ IPR at critical points,  $\braket{{\cal I}_{q,c}}$ with $N$. The top and bottom panels correspond to $q=1/4$ and $q=2$ respectively. The blue circles denote the data at the critical $W_c$ for a fixed $\xi$ (horizontal slice of the phase diagram) whereas the red circles denote the data  at the critical $\xi_c$ for a fixed $W$ (vertical slice). The dashed lines denote the fits to the functional forms in Eq.~\ref{eq:crit-fit}.}
\label{fig:ipr-crit-N}
\end{figure}

With the phase diagram of model established, we next focus on the nature of the Anderson transition.
In several recent studies~\cite{garciamata2017scaling,mace2019multifractal,roy2021fockspace,garciamata2022critical,sutradhar2022scaling,ghosh2025scaling,roy2025spectral}, finite-size scaling theory of the IPRs have emerged as an extremely useful tool to characterise localisation transitions, both in single-particle problems on high-dimensional graphs as well as the MBL transition on the Fock-space graph.
We will use a similar methodology here as well to understand the nature of the Anderson transition.

The key ingredient in the scaling theory is the scaling ansatz
\eq{
\braket{{\cal I}_q(N)} = \braket{{\cal I}_{q,c}(N)}\times {\cal F}_q\left(\frac{\cal S}{\Xi}\right)\,,
\label{eq:scaling-def}
}
where $\braket{{\cal I}_{q,c}}$ is the average $q^{\rm th}$ IPR at the critical point and ${\cal F}_q$ is the scaling function. In the argument of the scaling function, if ${\cal S} = N$, then $\Xi = \Lambda_q$ corresponds to a volume-scale on the graph whose behaviour with the parameters controls the critical scaling and yields the critical exponents characterising the transition.
On the other hand, if ${\cal S} = \ln N$, then $\Xi = \xi_q$ is a length-scale on the graph whose behaviour near the transition determines the relevant critical exponents. In current parlance, the former is often referred to as volumic scaling whereas the latter as linear scaling. 
To probe the complementary physics probed by the IPRs at small and large $q$, we will focus on $q=1/4$ and $q=2$ as representative values. 
In addition, we will denote by $\vartheta$, the difference $\delta W\equiv W-W_c$ when considering a horizontal slice of the phase diagram at a fixed $\xi$. 
Similarly, while considering a vertical slice of the phase diagram at a fixed $W$, we denote $\delta\xi\equiv \xi-\xi_c$ also by $\vartheta$.

Before delving into a discussion of the scaling theory, we discuss briefly the behaviour of the IPRs {\it at} the critical point where $q_{\ast,c}=1/2$. Therefore, for $q<1/2$, one expects $\tau_{q,c} = 2q - 1$, such that for $q=1/4$, we have $\braket{{\cal I}_{q=1/4,c}}\sim N^{1/2}$. This is in excellent agreement with the numerical result shown in Fig.~\ref{fig:ipr-crit-N} (upper panel).
The case of $q=2$ is somewhat more subtle. Theoretical results using the supersymmetric approach~\cite{fyodorov1991localisation,tikhonov2019critical} predict that $\braket{{\cal I}_{q,c}}\sim c_1 + c_2(\ln N)^{-1/2}$, effectively meaning asymptotic localisation in the thermodynamic limit. On the other hand, extending the idea that localised wavefunctions live on a few rare branches of the graph and exhibit multifractality on those branches at the critical point would suggest $\braket{{\cal I}_{q,c}}\sim (\ln N)^{-\sigma_q}$ with $0<\sigma_q<1$; this picture was found numerically in SWNs~\cite{garciamata2017scaling}.
For an unbiased result, we fit our data to a form 
\eq{
\braket{{\cal I}_{2,c}} = c_1 + c_2 (\ln N)^{-\sigma_2}\,,
\label{eq:crit-fit}
}
and the results are shown in Fig.~\ref{fig:ipr-crit-N} (lower panel).
We find that, in the best fit, $c_1\ll c_2$ and $\sigma_2\approx 0.5$.
This suggests that even if the wavefunctions are localised at the critical point, as the supersymmetric field theory predicts, they are very weakly so and for all practical purposes, they can be considered to be multifractal on a few rare branches of the graph.
What remains unambiguous is that the critical wavefunctions are strongly non-ergodic.

\begin{figure*}[t]
\includegraphics[width=\linewidth]{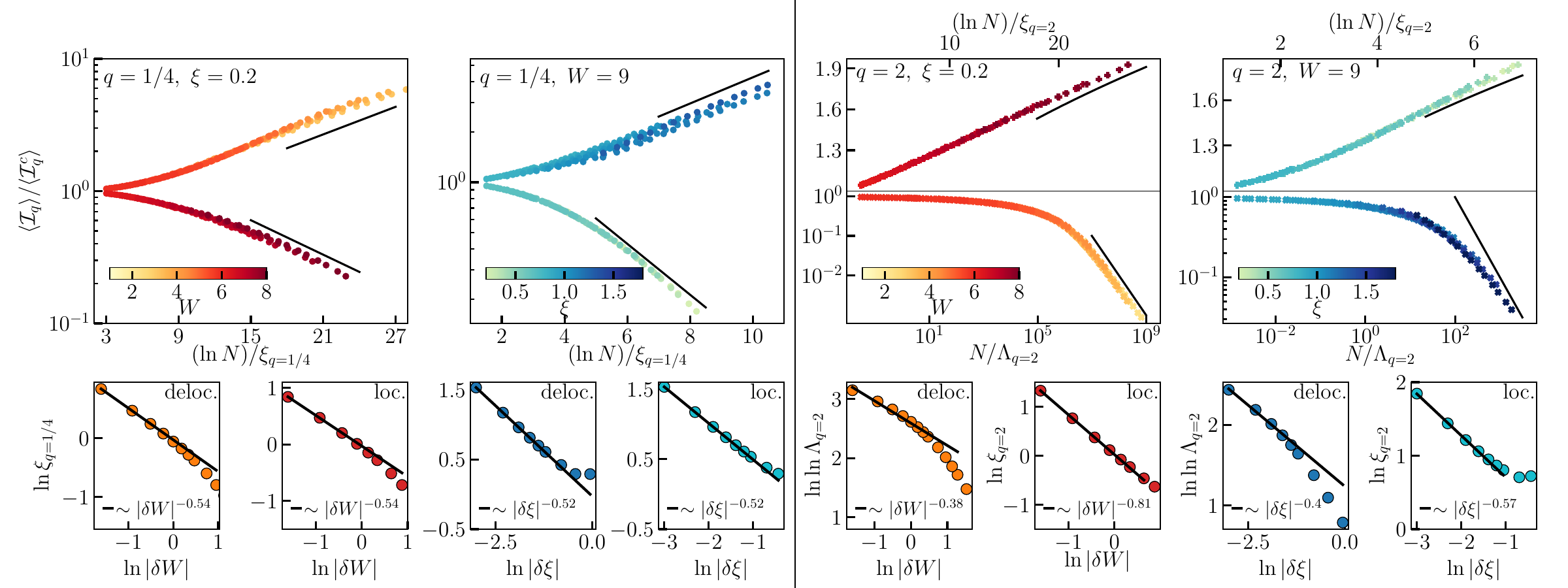}
    \caption{Results for the numerical scaling theory of the Anderson transitions in the ExpRRG model using the scaling of the IPRs. The left half of the figure corresponds to $q=1/4$ whereas the right half corresponds to $q=2$. The upper two panels in each half show the scaling collapse of $\braket{{\cal I}_q}/\braket{{\cal I}_{q,c}}$ when plotted as function of the scaling variable, $\ln N/\xi_q$ or $N/\Lambda_q$. The black solid lines denote the asymptotic forms of the scaling function as discussed in the main text. Amongst the two panels, the left one (with the red tones) corresponds to a horizontal slice of the phase diagram (varying $W$ at $\xi=0.2$ with the colourbar denoting the $W$ value) whereas the right ones correspond to vertical slice (varying $\xi$, represented by the colourbar, at $W=9$). The lower smaller panels show the how $\xi_q$ or $\Lambda_q$ diverges as the critical point is approached, $|\vartheta|\to 0$. 
    For $q=1/4$ (left half), the data shows the validity of the linear scaling in both the phases with a lengthscale $\xi_{q=1/4}$. The lower panels show that in both the phases, $\xi_{q=1/4}$ diverges as $|\vartheta|^{-\nu_{1/4}}$ with $\nu_{1/4}\approx 0.5$.
    The right half of the figure, corresponding to $q=2$, provides evidence for the fact that in the delocalised phase the data is well described by a volumic scaling (lower half of the two upper panels) whereas the localised phase is described by a linear scaling ansatz (upper half of the two upper panels). The lower panels show that as the transition is approached from the delocalised side, the volume-scale $\Lambda_2$ diverges as $\exp[b|\vartheta|^{-\nu_2^{\rm deloc}}]$ with $\nu_2^{\rm deloc}\approx 0.4$. On the other hand, the lengthscale $\xi_2$ diverges as a power-law on approaching the transition from the localised phase. The analysis presented has been done over a range of system sizes $N=2^i$ and $2^i \times 3/4$ with $i=8,9,\cdots,14$.
    In the scaling collapses, $W_c$ or $\xi_c$ was not used a fitting parameter and we used the estimate of the critical point from the level statistics data as the starting point. However, we also found that changing $W_c$ or $\xi_c$ by a small amount does not improve the quality of the collapse. 
    }
    \label{fig:ipr-scaling}
\end{figure*}

We now discuss the scaling theory of the transition in terms of the IPRs starting with $q=1/4$.
A point to note is that for any $q<1/2$, there must exists a $W_\ast(q)$ at a fixed $\xi$ (or a $\xi_\ast(q)$ at a fixed $W$) such that for $W>W_\ast(q)$ (or $\xi<\xi_\ast(q)$), $\tau_q = 0$. This is due to the fact that in the localised phase at some parameter value $(W,\xi)$, $\tau_q=0$ for $q>q_\ast(W,\xi)$ and $q_\ast$ decreases as one goes deeper into the localised phase. 
However, since our regime of interest is the vicinity of the transition (and where $q_\ast\to 1/2$) we will consider parameters for which $q_\ast$ is necessarily greater than 1/4.
As such for parameters considered in the localised, phase, we expect $\tau_{q=1/4} = 1/[4q_\ast(W,\xi)] - 1$.
\\

\noindent
{\it Scaling theory for $q=1/4$}:

The results for the scaling collapses of $\braket{{\cal I}_{q=1/4}}$ are presented in the left half of Fig.~\ref{fig:ipr-scaling}, which shows that the data is best described by a linear scaling function in both the phases with a correlation length that diverges as power law
\eq{
\xi_{q=1/4}^{\rm loc},\xi_{q=1/4}^{\rm deloc} \sim |\vartheta|^{-\nu_{1/4}}\,,
}
with $\nu_{1/4}\approx 0.5$ in both the phases. 

For such small values of $q$, in the localised phase, the IPR in the asymptotic limit of $\ln  N\gg\xi_q$ scales as $N^{-(q/q_\ast -1)}$. This implies that the aysmptotic of the scaling function is given by
\eq{
{\cal F}_{q=1/4}^{\rm lin/loc}(x) = 
    \begin{cases}
    \exp(-ax)\,&;~x\gg 1\\
    1\,&;~x\to 0
    \end{cases}\,,
}
with the constant $a>0$. 
This relation leads to
\eq{
q_{\ast,c}-q_{\ast} \sim \xi_{q=1/4}^{-1} \sim |\vartheta|^{\nu_{1/4}}\,,
}
which shows as the transition is approached from the localised side $q_\ast\to q_{\ast,c}=1/2$ as a square root of the distance from the critical point since $\nu_{1/4}\approx 0.5$.
From Eq.~\ref{eq:qstar-zeta-typ}, the above result is equivalent to the statement that the $\zeta_{\rm typ}$ approaches its critical value of $\zeta_{\rm typ}^c=1/2\ln(K-1)$ also as a square root of $|\vartheta|$,
\eq{
\zeta_{\rm typ} = \zeta_{\rm typ}^c - C |\vartheta|^{\nu_{1/4}}\,;~\nu_{1/4}\approx 0.5\,.
}
The lengthscale $\xi_{q=1/4}^{\rm loc}$ can have the following interpretation. In the localised phase, the typical correlation function decays exponentially with a lengthscale $\zeta_{\rm typ}$. As the it approaches the  critical value of  $\zeta_{\rm typ}^c=1/2\ln(K-1)$, the competition between exponential growth in $N_r$ with $r$ and the exponential decay of the typical correlations leads to the emergent strong multifractality at the critical point. 
The lengthscale $\xi_{q=1/4}^{\rm loc}$ simply encodes the difference via 
$\xi_{q=1/4}^{-1}\sim \zeta_{\rm typ}^c-\zeta_{\rm typ}$.

The results in Fig.~\ref{fig:ipr-scaling} (left half) show that in the delocalised phase as well, the scaling of $\braket{{\cal I}_{q=1/4}}$ is best described by a linear scaling. 
However, this is arguably paradoxical as it is not compatible with an asymptotically ergodic phase with $\braket{{\cal I}_q}\sim N^{-(q-1)}$.
The resolution to this is that indeed asymptotically at large $N$, the wavefunctions are ergodic and the linear scaling breaks down; however, in the critical regime there is a transient (in $N$), weak multifractality of the wavefunctions which manifests itself in $\tau_q$ deviating slightly from $q-1$ for $q<1$ and also equivalently, $q_\ast = 1-\varepsilon$ with $\varepsilon \to 0$ as $N\to \infty$.
To account for this, a deformed scaling function can be proposed~\cite{garciamata2022critical} which isn't purely of the linear form, but we will also discuss under what conditions does the linear scaling form emerge from it.
The deformed scaling function is given by
\eq{
{\cal F}^{\rm deloc}_{q}(N,\xi^{\rm deloc}_{q}) \sim \exp\left(bq\frac{\ln N}{\xi_{q}^{\rm deloc}}\right) + \left(\frac{N}{e^{\xi_q^{\rm deloc}}}\right)^{\tau_{q,c}-q+1}\,.
}
Using the above scaling function in Eq.~\ref{eq:scaling-def} we have
\eq{
\braket{{\cal I}_q} \sim N^{-(q/q_\ast-1)}+\frac{N^{-(q-1)}}{e^{\xi_q^{\rm deloc}q\left(\frac{1}{q_{\ast,c}}-1\right)}}\,,
\label{eq:Iq1/4-deloc}
}
where $q_\ast^{-1} = q_{\ast,c}^{-1}-b/\xi_q^{\rm deloc}$.
Since $q_\ast<1$ in the delocalied phase, the second term in Eq.~\ref{eq:Iq1/4-deloc} dominates in the limit of large $N$ and the ergodic scaling of $\braket{{\cal I}_q}\sim N^{-(q-1)}$ is recovered. 
On the other hand, the above equation also suggests that there is crossover scale
\eq{
N_{\ast,q} =\left[\exp({\xi_q^{\rm deloc}})\right]^{\frac{q_\ast}{1-q_\ast}}\,,
\label{eq:Nastq}
}
such that for $N\ll N_{\ast,q}$, the first term in Eq.~\ref{eq:Iq1/4-deloc} dominates. Near the critical point, since $q_\ast\to 1$, the right hand side in Eq.~\ref{eq:Nastq} can become arbitrarily large such that $N\ll N_{\ast,q}$ for a large range of $N$.
This leads to the linear scaling function being the appropriate one in the delocalised phase as well accompanied by a transient multifractality.
The picture that therefore emerges is that near the critical point there exists a lengthscale $\xi_q^{\rm deloc}$ for $q<1/2$ over which the wavefunctions show strong multifractality. As the critical point is approached, this lengthscale diverges and the global multifractality at such values of $q$, characteristic of the critical point, sets in. 
\\

\noindent
{\it Scaling theory for $q=2$}:

Let us now move to the case of $q=2$. 
In this case, we find that the delocalised phase is described by volumic scaling whereas the localised phase is characterised by a linear scaling of the data (see right half of Fig.~\ref{fig:ipr-scaling}). 
The data also shows that the volume-scale diverges on  approaching the transition from the delocalised phase as 
\eq{
\Lambda_2 \sim \exp[b |\vartheta|^{-\nu_2^{\rm deloc}}]\,,
}
with $\nu_2^{\rm deloc}\approx 0.4$.
Physically, this leads to the interpretation of the delocalised phase as a patchwork of non-ergodic volumes of volume-scale $\Lambda_2$. For $N \gg \Lambda_2$, the wavefunction can be viewed as an ergodic mosaic of such non-ergodic volumes such that on the global scales of the system size, the wavefunction is ergodic and the leading term is $\braket{{\cal I}_2}\sim N^{-1}$.
On the other hand, for $N\ll \Lambda_2$, a single patch of non-ergodic volume covers the entire system size and one sees the non-ergodicity, characteristic of the critical point, for such system sizes.
The asymptotic forms of the volumic scaling function in the delocalised phase can therefore be obtained as 
\eq{
{\cal F}^{\rm vol}_{q=2}(x)=
\begin{cases}
x^{-1}(\ln x)^{\sigma_2}\,;&x\gg1\\
1\,;&x\to0
\end{cases}\,,
}
leading to the asymptotic form of the IPR in the delocalised phase as 
\eq{
\braket{{\cal I}_2} \overset{N\gg\Lambda}{\sim} \frac{\Lambda}{N}\left(1-\sigma_2\frac{\ln \Lambda}{\ln N}\right)\,.
}

The scaling of $\braket{{\cal I}_2}$ in the localised phase, on the other hand, is best described by a linear scaling of the data characterised by a lengthscale $\xi_{q=2}$ which diverges as the critical point is approached as 
\eq{
\xi_{q=2}\sim |\vartheta|^{-\nu_2^{\rm loc}}\,.
}
\new{We find that the exponent $\nu_2^{\rm loc}$ is slightly different along the horizontal and vertical slices considered in the phase diagram, with the values being $\approx 0.8$ and $\approx 0.6$ respectively.
However, given that there exists a correlation length scale which diverges with the exponent $\nu_2^{\rm loc}$ upon approaching the transition from the localised phase, in the thermodynamic limit, we expect the two values to be the same. The magnitude of the discrepancy could therefore be seen as a lower bound to the systematic errors in its estimation within the limitations of our scaling analysis performed using finite-sized systems. 
}Nevertheless, the validity of the linear scaling in the localised phase remains robust.

Given that the $q=2$ IPR in the localised phase is independent of $N$, the asymptotic forms of the linear scaling function can be obtained as 
\eq{
{\cal F}^{\rm lin}_{q=2}(x)=
\begin{cases}
x^{\sigma_2}\,;&x\gg1\\
1\,;&x\to0
\end{cases}\,,
}
which is numerically confirmed in Fig.~\ref{fig:ipr-scaling}.
The aysmptotic form of the IPR in the localised phase is thence 
\eq{
\braket{{\cal I}_2}\overset{\ln N\gg\xi_2}{\sim} \frac{1}{\xi_{q=2}^{\sigma_2}}\,.
\label{eq:I2-asymp}
}

Physically, the divergent lengthscale $\xi_{q=2}$ leads to the following picture. As discussed earlier, a caricature of the localised phase is that the wavefuctions have overwhelmingly large support on a few rare branches of the graph. The average correlation function which decays exponentially with a scale $\zeta_{\rm avg}$ is sensitive to this. However, as argued in the previous section, at the critical point, this correlation length approaches a finite value $\zeta_{\rm avg}^{c} = 1/\ln(K-1)$, and this is sufficient to break the localisation as the exponential growth in the number of sites at distance $r$ can outweigh the exponential decay of the correlation.
The divergence of $\xi_2$ then controls how the correlation length $\zeta_{\rm avg}$  approaches its critical value of $\zeta^c_{\rm avg}$.
To see this formally, note that the asymptotic IPR in Eq.~\ref{eq:I2-asymp} must be the same as that in the first line of Eq.~\ref{eq:I2-corr}. Equating the two and assuming $\xi_{q=2}\gg 1$ leads to the relation 
\eq{
\zeta^c_{\rm avg}-\zeta_{\rm avg} \sim \xi_{q=2}^{-\sigma_2} \sim |\vartheta|^{\sigma_2\nu_2^{\rm loc}}\,,
}
which explicitly shows that the approach of $\zeta_{\rm avg}$ to the critical value of $\zeta^c_{\rm avg}$ is controlled by the exponents $\sigma_2$ and $\nu_2^{\rm loc}$. 
Previous studies have suggested numerically that $\nu_{2}^{\rm loc}=1$~\cite{garciamata2017scaling,garciamata2022critical} and while along the horizontal slide our result of $\approx 0.8$ is not too different from that, 
along the vertical cut our result of $\approx 0.6$ 
\new{is more different. However, as mentioned earlier, we expect that the two exponents would become equal if the scaling analysis could be performed with significantly larger systems sizes.}

\new{
The upshot of the above scaling analysis is that asymptotically, in the $N\to\infty$ limit, the critical point is described by volumic and linear scaling laws on approaching from the delocalised and localised sides respectively. 
The form of the divergence of the volume-scales and lengthscales on approaching the transition from the delocalised and localised phases indicates a Kosterlitz-Thouless-like scaling for the transition.
However, for small $q$, there is a parametrically large transient scale in $N$ (see Eq.~\ref{eq:Nastq}) below which the delocalised side is also governed by linear scaling.
}

\subsection{Self-consistent theory \label{sec:selfconsisten}}

We now turn towards an analytic, albeit approximate, derivation of the Anderson transitions using the local Green's function $G_{xx}(\omega)$, defined in Eq.~\ref{eq:Gxx}. 
However, instead of focussing on $G_{xx}(\omega)$ directly, we will find it more convenient to work with the local self-energy, $\Sigma_x(\omega)$ defined via 
\eq{
G_{xx}(\omega) = [\omega^+-\epsilon_x-\Sigma_x(\omega)]^{-1}\,,
}
where $\Sigma_x(\omega) = {\cal R}_x(\omega)-i\Delta_x(\omega)$ with ${\cal R}_x$ and $\Delta_x$ respectively being the real and imaginary parts of the self-energy.
Of interest is the latter as it encodes physically the rate of loss of probability from site $x$ into eigenstates at energy $\omega$.
In a delocalised phase, $\Delta_x(\omega)$ is finite with unit probability in the thermodynamic limit, whereas in the localised phase it is proportional to the line broadening and hence vanishes $\propto \eta$.
Analogous to the LDoS or equivalently the imaginary part of the local Green's function (see Eq.~\ref{eq:Dx-def} and Eq.~\ref{eq:Dx-G}), $\Delta_x(\omega)$ therefore constitutes a probabilistic order parameter for the localisation transition.
Indeed, within a Feenberg renormalised perturbation series (RPS), the self-energy can be expressed as~\cite{Economoubook,feenberg1948note,watson1957multiple,abou-chacra1973self,logan1985anderson,logan1987dephasing}
\eq{
\Sigma_x(\omega) &= \sum_y |t_{xy}|^2G_{yy}(\omega)+\cdots\nonumber\\
&= \sum_y \frac{|t_{xy}|^2}{\omega^+-\epsilon_y-\Sigma_y(\omega)}+\cdots\,,\label{eq:Delta-RPS}
}
where first line makes the analogous behaviour between the local Green's function and the local self-energy explicit. 
From Eq.~\ref{eq:Delta-RPS}, the RPS for the real and imaginary parts can be written explicitly to leading order as 
\begin{subequations}\label{eq:self-consistent}
\begin{align}
\mathcal{R}_x(\omega) &=\!\sum_{y} \dfrac{|{ t_{xy}}|^2(\omega-\eps_{y} -\mathcal{R}_y(\omega))}{(\omega-\eps_{y} -\mathcal{R}_y(\omega))^2+(\eta+\Delta_y(\omega))^2}\,, \label{eq:reS} \\
\Delta_x(\omega)&=\!\sum_{ y} \dfrac{|{ t_{xy}}|^2(\eta+\Delta_y(\omega))}{(\omega-\eps_{y} -\mathcal{R}_y(\omega))^2+(\eta+\Delta_y(\omega))^2}\,. \label{eq:imS}
\end{align}
\end{subequations}
In the following, we will analyse the above equations using a self-consistent method numerically, analogous to a belief propagation method, and a self-consistent mean-field approximation analytically~\cite{abou-chacra1973self,logan1985anderson,logan1987dephasing,logan1987dipolar,logan2019many,roy2020fock,roy2019self,DuthieSelfConPRB2021}.

\begin{figure}
    \centering
    \includegraphics[width=\linewidth]{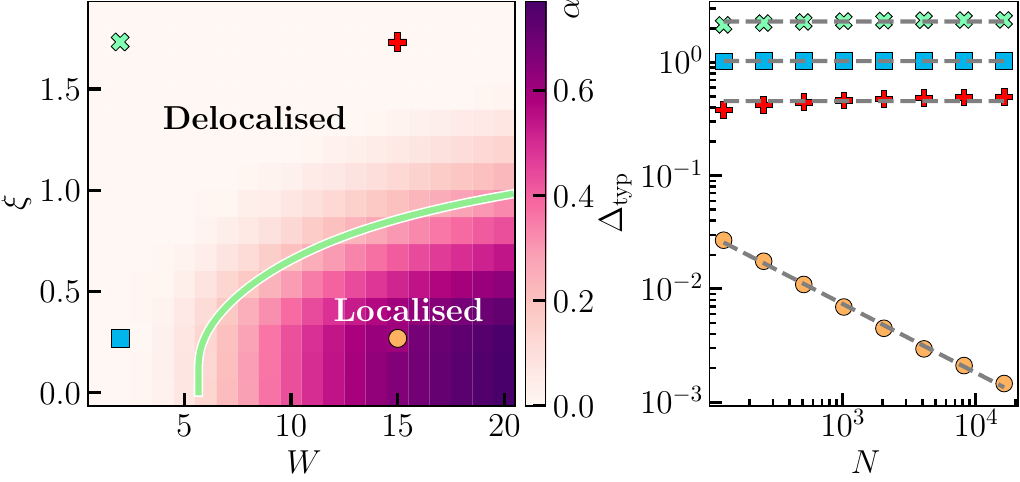}
    \caption{The phase diagram of the ExpRRG model as obtained from the scaling of the imaginary part of the self-energy, $\Delta_{\rm typ}\sim N^{-\alpha}$. The left panel shows $\alpha$ as a heatmap with lighter colours corresponding to $\alpha\approx 0$ denoting the delocalised phase whereas the darker colours denoting $\alpha\lessapprox 1$ denoting the localised phase. The green line denotes the analytical result from the self-consistent mean-field theory (Eq.~\ref{eq:Wc-mf-loc}). The right panel shows representative data for $\Delta_{\rm typ}$ as a function of $N$ at four points in the parameter space, denoted by the different markers.}
    \label{fig:recursive}
\end{figure}

In the former, we numerically solve the system of equations in Eq.~\ref{eq:self-consistent} iteratively for individual realisations of the graph and disorder potentials.
Specifically, we start with some initial seed values of $\{{\cal R}_x(\omega),\Delta_x(\omega)\}$ for all $x$ and update them using Eq.~\ref{eq:self-consistent}.
This recursive process is continued until the values  converge to a fixed-point solution, $\{{\cal R}_x^{\rm fp}(\omega),\Delta_x^{\rm fp}(\omega)\}$, which satisfies Eq.~\ref{eq:self-consistent} for all $x$.
The typical value of $\Delta_x(\omega)$ is then computed as 
\eq{
\Delta_{\rm typ}(\omega) = \exp\braket{\ln \Delta_x^{\rm fp}(\omega)}\,,
}
where the $\braket{\cdots}$ denotes the average over realisations of graphs and disordered potentials, as well as sites. 
The delocalised phase is signified by $\Delta_{\rm typ}(\omega)\sim N^0$ whereas in the localised phase   $\Delta_{\rm typ}(\omega)\sim\eta$.
In our calculations, we consider $\eta\sim N^{-1}$ and focus entirely on $\omega=0$.
Numerically, we fit the data to a form 
\eq{
\Delta_{\rm typ}(\omega)\sim N^{-\alpha}\,,
}
and show $\alpha$ has a heatmap in the $W$-$\xi$ plane in Fig.~\ref{fig:recursive}
The phase boundary is indicated by $\alpha$ crossing over sharply from $\approx 0$ in the delocalised phase to $\lessapprox 1$ in the localised phase. 
With the phase diagram, using the RPS for $\Sigma_x(\omega)$ established numerically, we next analyse Eq.~\ref{eq:self-consistent} analytically at a mean-field level.

To set up the mean-field theory, we replace $\{{\cal R}_x(\omega),\Delta_x(\omega)\}$ in the right-hand side of Eq.~\ref{eq:self-consistent} by its typical value. In addition, since our focus is on $\omega=0$, statistically speaking, ${\cal R}_x(\omega)$ is expected typically to be zero due to a $\omega \leftrightarrow -\omega$ symmetry in the spectrum on an average.
We therefore have
\eq{
\Delta_x = \sum_{y}\frac{|t_{xy}|^2 (\eta +\Delta_{\rm typ})}{\epsilon_y^2 + (\eta + \Delta_{\rm typ})^2}\,,
\label{eq:Delta-sc}
}
where we have set $\omega=0$ and suppressed it in the notation for brevity. 
A distribution of $\Delta$, denoted by $P_{\Delta}(\Delta,\Delta_{\rm typ})$ is obtained by transforming the distribution of $\{\eps_y\}$ using the relation between the two in Eq.~\ref{eq:Delta-sc}.
Self-consistency is then imposed by equating the $\Delta_{\rm typ}$ obtained from the distribution to the one that enters the distribution parametrically,
\eq{
\ln \Delta_{\rm typ} = \int d\Delta~P_{\Delta}(\Delta,\Delta_{\rm typ})\ln\Delta\,.
\label{eq:Delta-sc-mf}
}
In the following, we analyse these equations separately in the localised and delocalised phases starting with the former.

\subsubsection{Localised phase}
In the localised phase, $\Delta_x$ is expected to be typically $\sim \eta$ and hence the quantity which admits a well-defined thermodynamic limit is $\delta_x = \Delta_x/\eta$. 
Moreover, the second term in the denominators in Eq.~\ref{eq:Delta-sc-mf} can be neglected in favour of the $\epsilon_y^2$ as the former is only $O(\eta^2)$ whereas the latter is $O(W^2)$ and hence $O(1)$. We therefore have
\eq{
\delta_x = (1+\delta_{\rm typ})\sum_y \frac{|t_{xy}|^2}{\epsilon_y^2}\,,
}
and thus $\delta_x$ is given by a sum of independent random numbers $\kappa|t_{xy}|^2/\eps_y^2$ where $\kappa =  1+\delta_{\rm typ}$
Since $\eps_y\sim {\cal N}(0,W)$ is a normally distributed random number, $\kappa|t_{xy}|^2/\eps_y^2$ has L\'evy distribution\footnote{The probability distribution function of a L\'evy distribution is given by $\levy(x;\mu,c)=\sqrt{\frac{c}{2\pi}}(x-\mu)^{-3/2}\exp\left(-\frac{c}{2(x-\mu)}\right)$.}, 
\eq{
\frac{\kappa|t_{xy}|^2}{\eps_y^2}\sim \levy\left(0,\frac{\kappa|t_{xy}|^2}{W^2}\right)\,.
}
Since, the $\levy$ distribution is a stable distribution, a sum of $\levy$ distributed random numbers is also $\levy$ distributed.
We therefore have $\delta_x\sim \levy\left(0,c\right)$ where
\eq{
c = \left(\sum_y \sqrt{\frac{\kappa |t_{xy}|^2}{W^2}}\right)^2 = \frac{\kappa K^2t^2}{W^2[1-(K-1)e^{-1/\xi}]^2}\,.
}
The typical value of a random number distributed as $\levy(0,c)$ is given by $2ce^{\gamma}$ where $\gamma$ is the Euler-Mascheroni constant. This yields
\eq{
\delta_{\rm typ} = \frac{2(1+\delta_{\rm typ}) K^2t^2e^{\gamma}}{W^2[1-(K-1)e^{-1/\xi}]^2}\,,
}
which establishes the self-consistent relation for $\delta_{\rm typ}$ in the localised phase. 
The above equation can be rearranged to give
\eq{
\delta_{\rm typ}=\frac{\Gamma}{1-\Gamma}\,;~\Gamma = \frac{2 K^2t^2e^{\gamma}}{W^2[1-(K-1)e^{-1/\xi}]^2}\,.
}
For the self-consistent solution to valid, we require $\delta_{\rm typ}>0$ or equivalently $\Gamma<1$. In order words, the self-consistency of the localised phase breaks down at $\Gamma=1$ which provides an estimate of the critical disorder strength as 
\eq{
 W_c(\xi)=\frac{\sqrt{2e^\gamma}Kt}{1-(K-1)e^{-1/\xi}}\,.
 \label{eq:Wc-mf-loc}
}
The locus of the critical points given by the above expression is shown in Fig.~\ref{fig:pd-exprrg} as well as in Fig.~\ref{fig:recursive} and it is indeed in good agreement with the complimentary numerical results.

\subsubsection{Delocalised phase}
In the delocalised phase, $\Delta_{\rm typ}$ is typically expected to be $O(1)$ in the thermodynamic limit. Therefore the limit of $\eta\to0$ can be taken from outset in Eq.~\ref{eq:Delta-RPS}, such that
\eq{
\Delta_x = \Delta_{\rm typ}\sum_{y}\frac{|t_{xy}|^2 }{\epsilon_y^2 + \Delta_{\rm typ}^2}\,.
}
Self-consistency implies that 
\eq{
\left\langle\ln \sum_{y}\frac{|t_{xy}|^2 }{\epsilon_y^2 + \Delta_{\rm typ}^2}\right\rangle = 0\,.
}
However, since our main focus is on approaching the the Anderson transition from the delocalised side, we expect that in this regime $\Delta_{\rm typ} \ll 1$ and in fact approaches zero as the transition is approached. As such, it is justified to neglect it in favour of $\eps_y^2$ in the denominator above. 
The condition for the critical point then becomes
\eq{
\left\langle\ln \sum_{y}\frac{|t_{xy}|^2 }{\epsilon_y^2}\right\rangle = 0 \Rightarrow \chi_{\rm typ}=1\,,
}
where $\chi = \sum_{y}{|t_{xy}|^2 }/{\epsilon_y^2}$ is again a sum of $\levy$ distributed random numbers and hence is itself $\levy$ distributed, $\chi\sim\levy(0,c')$ with
\eq{
c'=\frac{ K^2t^2}{W^2[1-(K-1)e^{-1/\xi}]^2}\,.
}
This implies that $\chi_{\rm typ}=2c'e^{\gamma}$, setting which to unity again yields
\eq{
 W_c(\xi)=\frac{\sqrt{2e^\gamma}Kt}{1-(K-1)e^{-1/\xi}}\,.
 \label{eq:Wc-mf-deloc}
}
Crucially, the critical disorder is identical to that obtained from the self-consistent theory in the localised phase in Eq.~\ref{eq:Wc-mf-loc}.
This implies that the self-consistency imposed separately in either phase breaks down exactly at the same critical disorder, which shows that the theory is internally consistent.
This concludes our derivation of the critical disorder within the self-consistent, mean-field approximation.

\new{Note that in the limit of $\xi\to 0$, the result for the critical disorder in Eq.~\ref{eq:Wc-mf-loc} or Eq.~\ref{eq:Wc-mf-deloc} scales linearly with $K$ as expected for short-ranged Anderson models.
However, the upper limit approximation~\cite{anderson1958absence,abou-chacra1973self} which probes the large $\Delta$'s in the tails of its distribution or considering the maximum $\Delta$ over all possible forward scattering paths~\cite{pietracaprina2016forward} leads to multiplicative correction of $\ln K$ to the $W_c$.
In Appendix~\ref{app:ula} we show that within the upper limit approximation, $W_c$ gets a multiplicative correction which scales as $\ln K$ for $\xi\ll 1$ whereas at intermediate $\xi$, it scales faster than $\ln K$. 
This again reflects the complex interplay between $K$ and $\xi$ in establishing the localisation phase diagram of the model.
}

{\let\MakeTextUppercase\relax
 \section{ExpRRG-RH Model} \label{sec:exprrg-rh}
}

In this section, we discuss the ExpRRG-RH model described by random hoppings on the random graph distributed according to Eq.~\ref{eq:exprrg-rh-dist}.
Note that the typical or average hopping amplitude in this case is the same as that of the ExpRRG model, $\braket{|t_r|} = e^{-(r-1)/\xi}$.
We therefore expect that the global phase diagram of the model to be qualitatively similar to that of the ExpRRG model, particularly away from the critical line. 
However, given the distribution of the random hoppings have a finite weight on vanishingly small values, it leads to the possibility of having weak links on the graph characterised by arbitrarily small hoppings. 

The pertinent question that arises therefore is, does the ExpRRG-RH model continue to have a direct Anderson transition between the delocalised and localised phases as in the ExpRRG model, or does the critical line open up into a robust multifractal phase interspersing the delocalised and localised phases.
The second possibility is indeed realised robustly in randomly-weighted Erd\H{o}s-R\'enyi graphs~\cite{CugliandoloPRB2024}.
Addressing the question would therefore amount to understanding the fate of multifractality in randomly-weighted Erd\H{o}s-R\'enyi graphs but in their spatially-structured, power-law banded versions.

Two limiting regions of the phase diagram are straightforward to understand. At weak disorder, (small $W$), and larger values of $\xi$ (longer-ranged hoppings), the model is anyway expected to be in a delocalised phase. Similarly at large $W$ and small $\xi$, the model is expected to be in a localised phase. 
The interesting case is either with both $W,\xi$ small or with both $W,\xi$ large as in this cases the two parameters potentially compete against each other, thereby opening up the possibility of a multifractal phase. 
We therefore focus on two slices of the phase diagram, namely at fixed small $\xi=0.2$ with varying $W$ and at fixed large $W=8$ and varying $\xi$. Note that these are the similar horizontal and vertical cuts in the phase diagram considered for the ExpRRG model earlier. 

From a numerical perspective we will focus on the level statistics \eqref{eq:lsr}, the generalised IPRs \eqref{eq:ipr-gen-q} and the typical correlation function \eqref{eq:Ctyp} as $\zeta_{\rm typ}$ obtained from the latter is directly related to $q_\ast$. 
It will be useful to recall also the signature of a robust multifractal phase in terms these quantities -- such a phase, if present, will manifest itself as an extended regime in parameter space where $1/2<q_\ast<1$ and $0<\tau_2<1$ in the thermodynamic limit with the inequalities being strict.

\begin{figure}
    \centering
\includegraphics[width=\linewidth]{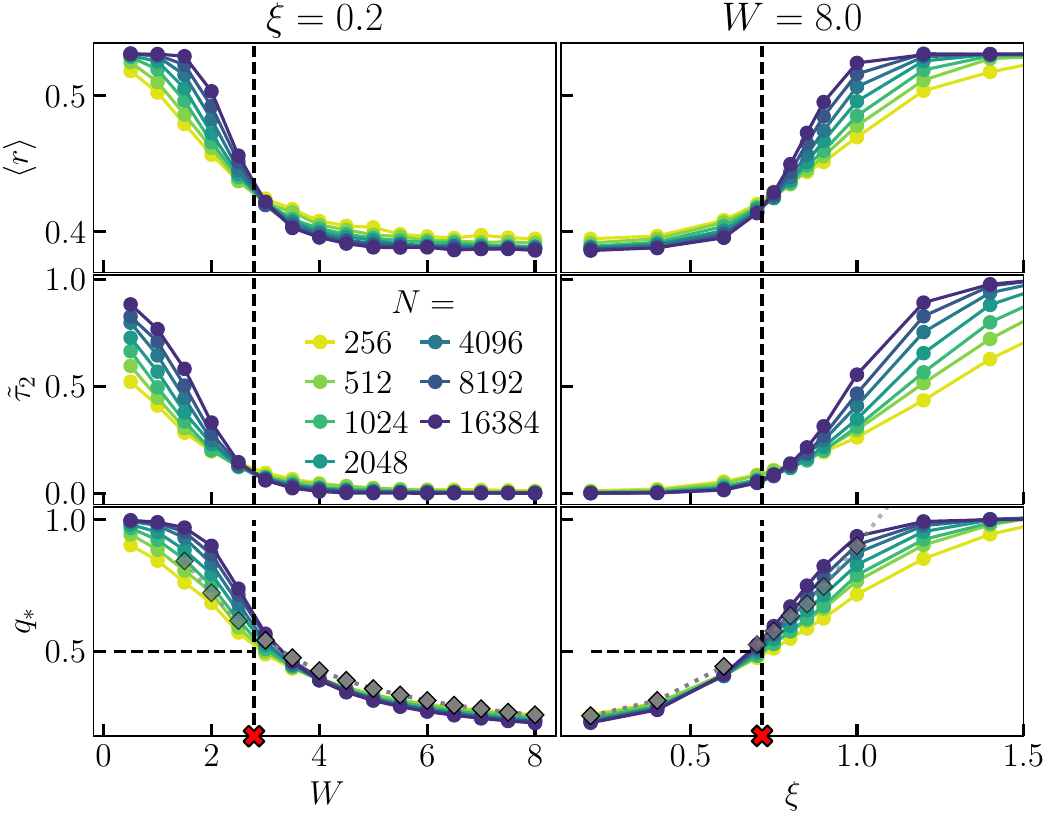}
    \caption{Numerical results for the ExpRRG-RH model along a horizontal (left) and a vertical cut (right) in the $W-\xi$ plane. Top panel shows the mean of the level spacing ratio, $\langle r\rangle$ of the data for different $N$ along the cuts. Along the $\xi=0.2$ cut, a transition from delocalised to localised phase is observed above a critical $W_c$. On the other hand, a transition from the localised to delocalised phase is visible when $\xi$ is increased along a vertical cut at $W=8$  at some $\xi_c$. The middle panel depicts $\tilde{\tau}_2$, as obtained from Eq.~\ref{eq:flowing-fractal-exponent}, along the cuts for different $N$. The two sides of the dashed lines either approach $0$ or $1$ with increase in $N$, depending on whether the phase is localised or delocalised respectively. Similarly, the bottom panel illustrates how $q_\ast$ varies as a function of the varying parameter ($W$ or $\xi$) for different $N$. In the localised phase, $q_*$ values comply with Eq.~\ref{eq:qstar-zeta-typ}, where dark gray squares represent $\zeta_{\rm typ}\ln (K-1)$. The convergence of $q_\ast$ at less than $1/2$ (for large $W$ along the horizontal cut and  small $\xi$ along the vertical cut) signifies the absence of an intermediate non-ergodic multifractal phase. The red crosses denote the critical points, obtained from the finite size scaling analysis of $\langle r \rangle$. }
    \label{fig:ExpRRG-RH}
\end{figure}

Based on these, the numerical results shown in Fig.~\ref{fig:ExpRRG-RH}, suggests that there does {\it not} exist a robust multifractal phase in the ExpRRG-RH model and the Anderson transition a direct transition between delocalised and localised phases.
In Fig.~\ref{fig:ExpRRG-RH}, the left column corresponds to the horizontal slice (fixed $\xi$, varying $W$) whereas the right column corresponds to the vertical slice (fixed $W$, varying $\xi$).
The top row shows the results for the level statistics, which show a clear crossing for different $N$ as the respective parameters are varied; however this only indicates a transition from an ergodic phase to a non-ergodic one (and not necessarily a localised one).

More importantly, as shown in the middle row, $\tilde{\tau}_2$ also seemingly shows a crossing for different $N$. At large $W$ for $\xi=0.2$ and at small $\xi$ for $W=8$, the data systematically goes down with $N$ towards zero, indicating a localised phase. On the other hand, at small $W$ for $\xi=0.2$ and at large $\xi$ for $W=8$, $\tilde{\tau}_2$ grows with $N$ towards 1. Crucially, within our numerical results, we do not find an extended parameter regime where $\tilde{\tau}_2$ is converged with $N$ at a value which is strictly $>0$ and $<1$. This indicates the absence of a multifractal phase. 

This is further corroborated in the bottom where we plot both $q_\ast$ as well as $\zeta_{\rm typ}\ln(K-1)$ as a function of the parameter. Again, we find that $q_\ast$ and $\zeta_{\rm typ}\ln(K-1)$ are very close to each other at large $W$ and small $\xi$ indicating a localised phase. More tellingly, we again do not find an extended parameter regime where $q_\ast$ is converged with $N$ at a value which is strictly $<1/2$. In fact, in both the panels, we find that once $q_\ast$ reaches $1/2$, it in fact starts growing towards 1 indicating a transition to the delocalised phase. 

All of these results collectively suggest that the ExpRRG-RH model does not host a robust multifractal phase and the multifractality in this model is, perhaps, limited only to the critical lines just like in the ExpRRG model.
In the following, we present an extremely heuristic argument for why the long-ranged nature of the hopping in the ExpRRG-RH completely destabilises the multifractality, found for example in the randomly-weighted Erd\H{o}s-R\'enyi graphs and leads to direct delocalisation from a localised phase. 

\begin{figure}
    \centering
\includegraphics[width=\linewidth]{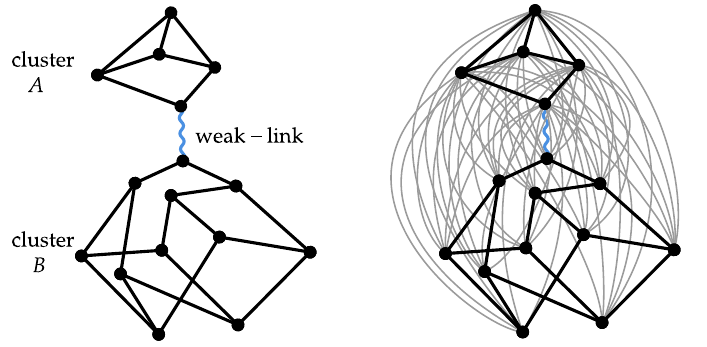}
    \caption{Left: Weak link in a randomly-weighted Erd\H{o}s-R\'enyi graph two clusters, $A$ and $B$, are connected by a single link on which the hopping amplitude is vanishingly small with a finite probability. Right: In the ExpRRG-RH model every site in $A$ is connected to every site in $B$ which naturally means that the probability of the clusters remaining weakly connected is exponentially suppressed in the number of sites in the graph. }
    \label{fig:ExpRRG-RH-mech}
\end{figure}

To present the argument, let us first recall that the main underlying mechanism of multifractality in randomly-weighted Erd\H{o}s-R\'enyi graphs (with finite average connectivity) was the presence of weak links~\cite{CugliandoloPRB2024}. For instance, consider the situation shown in the left panel of Fig.~\ref{fig:ExpRRG-RH-mech} where there exist two clusters of sites in a realisation of an Erd\H{o}s-R\'enyi graphs, and one of the clusters, say cluster $A$ has only $N^{D_A}$ sites with $D_A<1$.
While each of the clusters can be quite ergodic in itself, there exists that a finite probability of the two clusters being connected by a single edge on the graph -- we will refer to these sites connecting the two clusters as the bridge sites. More importantly, the fact that the hoppings are random with a finite probability of a vanishingly small hopping implies there exists a finite probability of the link between the bridge sites being a weak link. 
In such a situation there will be $N^{D_A}$ eigenstates of the model which will be majorly supported only within cluster $A$. Even if they are ergodic within the cluster, there IPR will scale as ${\cal I}_q\sim N^{-D_A(q-1)}$ indicating multifractality as $D_A<1$.

Contrast this to the case of the ExpRRG-RH model where the long-ranged hoppings would lead to hopping from all sites in cluster $A$ to all sites in cluster $B$, albeit with different magnitudes. For the weak-link mechanism to survive, one therefore requires all the hoppings between cluster $A$ and cluster $B$ are necessarily weak -- this is because even one of the hoppings becoming resonant is sufficient to kill the multifractality as it provides a pathway for the eigenstate to delocalised ergodically over both the clusters. 
Since the number of hoppings between the two clusters is $\sim N^{D_A+D_B}$, it diverges in the $N\to\infty$ limit. It therefore stands to reason that the probability of all the links being weak is exponentially suppressed in $N$ leading to the absence of multifractality in the $N\gg 1$ limit.

To make this slightly formal, let us denote the probability of a bond between two sites at distance $r$ from each other being resonant be given by $p_{\rm res}(r)$.
Given that our skeleton graph is an RRG, it is natural to assume that each cluster forms an RRG with its own effective diameter, say $r_A$ and $r_B$. The probability that clusters $A$ and $B$ are weakly connected is then given by
\eq{
p_{\text{weak-link}}^{AB} = \prod_{r_1=0}^{r_A}\prod_{r_2=0}^{r_B} [1-p_{\rm res}(r_1+r_2)]^{K^2(K-1)^{r_1+r_2}}\,.
\label{eq:p-weak-link}
}
The above expression can be understood simply as the following.
The probability that a given pair of sites, one in cluster $A$ at distance $r_1$ from the bridge and the other in cluster $B$ at distance $r_2$ from the bridge, is not resonant is $1-p_{\rm res}(r_1+r_2)$. At the same time the total number of such pairs of sites is $N_{r_1}N_{r_2} = K(K-1)^{r_1-1}\times K(K-1)^{r_2-1}$. The probability that clusters $A$ and $B$ are therefore weakly connected is given by the product over all such probabilities.
In the thermodynamic limit, both $r_A,r_B \to \infty$. This effectively means that Eq.~\ref{eq:p-weak-link} is simply a product of infinite number of probabilities each of which is necessarily less than 1, and therefore goes to zero. 
This provides one heuristic argument for why the weak-link phenomenon and therefore the multifractality does not survive in the $N\to\infty$ limit for the ExpRRG-RH model.

\section{Conclusion \label{sec:summary}}

We conclude with a brief summary of the work and some remarks by way of future outlook.
In this work, we introduced a new class of Anderson models on random graphs by endowing them with spatially structured hopping, which corresponds to a power-law hopping generalisation of the conventional Anderson models on random regular graphs.
This was implemented via a modulation of the hopping amplitudes that decays exponentially with the distance between sites on the graph, with a characteristic lengthscale $\xi$.
Using a combination of numerical results from exact diagonalisation and analytical results from a self-consistent renormalised perturbation theory, we charted the localisation phase diagram in the parameter space of $\xi$ and the onsite disorder strength $W$.
We found that with increasing $\xi$, the critical disorder strength increases.
However, there exists a value of $\xi$ above which the delocalised phase persists even for arbitrarily strong disorder.
In addition, all our results point towards a direct Anderson transition between the delocalised and localised phases, with no intervening multifractal phase.
Based on a scaling theory of the inverse participation ratios, we find that the transition is consistent with a Kosterlitz–Thouless-like scenario and admits a two-parameter scaling, in line with the generic situation in high-dimensional graphs~\cite{garciamata2017scaling,garciamata2022critical}.

A natural question to ask is what the dynamical phase diagram of this class of models looks like.
It has been shown in earlier works that the dynamical response of disordered systems on high-dimensional graphs, such as that encoded in the dynamics of the return probability and wavepacket spreading, can exhibit a much richer structure than the corresponding localisation phase diagrams~\cite{detomasi2019survival,detomasi2020subdiffusion,khaymovich2021dynamical}.
In particular, it will be interesting to study the fate of the subdiffusive behaviour in the delocalised phase in the vicinity of the Anderson transition, observed for nearest-neighbour models on the random regular graph~\cite{detomasi2020subdiffusion}, in the presence of the long-ranged hoppings considered in this work.

Relatedly, it will also be of interest to study the possibly anomalous transport properties in such models using methods such as non-equilibrium Green’s functions, and to uncover the local profile of currents therein.
This may shed light on the microscopic origins of the large-deviation forms~\cite{biroli2022critical,biroli2024largedeviation} observed in non-local Green’s functions in the presence of long-ranged hopping.

An additional motivation for studying localisation problems on high-dimensional graphs with long-ranged spatial structure is to interpret them as proxies for the many-body localisation problem on the Fock-space graph.
Specifically, one aspect that this construction, or a variant thereof, can capture is the non-perturbative effects of long-ranged rare resonances in the many-body localisation problem~\cite{morningstar2021avalanches,crowley2022constructive,garratt2021resonances,garratt2022resonant}.
However, as our analysis shows, neither the ExpRRG nor the ExpRRG-RH model hosts a multifractal phase, which is a characteristic feature of many-body localised eigenstates on the Fock-space graph~\cite{deluca2013ergodicity,mace2019multifractal,roy2021fockspace}.
With this underlying motivation, our construction of random graphs with spatial structure opens the possibility of studying several other, qualitatively different, generalisations of the model that may be more faithful to many-body localisation physics.
In particular, it will be interesting to study a version of the random-hopping model in which the matrix elements are drawn from Lévy distributions.
Relatedly, one could also consider a variant in which the hopping amplitude between two sites at distance $r$ is set to an $O(1)$ value with a probability that scales as a power law, $r^{-\alpha}$, and is zero otherwise.
One may expect that such constructions better mimic the physics of long-ranged rare resonances, wherein the power-law nature of the matrix elements destabilises the localised phase and drives it towards a multifractal regime.

\acknowledgments
We thank A. Dhar for useful discussions and comments on the manuscript. SR acknowledges support from SERB-DST, Government of India, under Grant No. SRG/2023/000858 and from a Max Planck Partner Group grant between ICTS-TIFR, Bengaluru and MPIPKS, Dresden.
The authors acknowledge support of the Department of Atomic Energy, Government of India, under project no. RTI4001.

\appendix
\new{
\section{Corrections to mean-field solution from the Upper Limit Approximation \label{app:ula}}
In this appendix, we 
present results for the critical disorder within
the `upper limit approximation' (ULA)~\cite{anderson1958absence,abou-chacra1973self} 
for the ExpRRG model.
The ULA is equivalent to estimating the critical disorder by considering the maximum $\Delta$'s over the available processes~\cite{pietracaprina2016forward}. 
For short-ranged Anderson models on trees or RRGs, the ULA leads to a critial disorder $W_c\sim K\ln K$, {\it i.e.} an additional multiplicative correction of $\ln K$ to the mean-field result.
As we show below, for the ExpRRG model, the analogous correction is a non-trivial function of $K$ and $\xi$.

To begin with, recall that in the localised phase at $\omega=0$, the imaginary part of the self-energy in Eq.~\ref{eq:imS} becomes
 \begin{align}
     \Delta_x&=\!\sum_{ y} \dfrac{|{ t_{xy}}|^2(\eta+\Delta_y)}{\eps_{y} ^2}\,.\label{eq:upa-delta}
 \end{align}
The ULA involves the probing the $\Delta_x$'s in the tails of the distribution so that one estimates an upper limit to the critical disorder.
It is therefore useful to study the Laplace transform $F(s)$ of $P_\Delta(\Delta)$ for $s\ll 1$. 
The Laplace transform can be expressed as 
\begin{align}
    F(s)=\prod_{y\neq x} \left[ \int d\eps_y P(\eps_y)F(s\rho_{xy}) \exp\left(-\eta s \rho_{xy}\right)\right] \,. \label{eq:laplace-delta}
\end{align}
with $\rho_{xy}=\abs{t_{xy}}^2/\eps_y^2$ and $P(\epsilon)$ is the distribution of the onsite potentials (Gaussian with zero mean and standard deviation $W$).
Assuming $P_\Delta(\Delta) \overset{\Delta\gg1}{\sim} \Delta^{-1-\beta}$, $F(s)$ at small $s$ can be approximated as $ F(s)\overset{s\ll1}{\approx}1-s^\beta$. 
The fat tailed nature of the $P_\Delta$ forces $0<\beta<1/2$.

Using this approximation in Eq.~\ref{eq:laplace-delta} and equating coefficients of $s^\beta$ leads to the following condition for the critical point
\begin{align}
    \sum_{y\neq x} \abs{t_{xy}}^{2\beta} \int P(\eps) \eps^{-2\beta}d\eps=1\,.
\end{align}
which for the ExpRRG model becomes
\begin{align}
    \frac{\Gamma(1/2-\beta)}{\sqrt{\pi} (2W^2)^\beta}\dfrac{K}{1-(K-1)e^{-2 \beta /\xi}} =1\,. \label{eq:ula-condition}
\end{align}
with $2\beta>\xi \ln (K-1)$. 
The expression in Eq.~\ref{eq:ula-condition} can be inverted to yield 
\eq{
W_\ast(\beta,\xi,K) = \frac{1}{\sqrt{2}}\left[\frac{K}{\sqrt{\pi}}\dfrac{ \Gamma(1/2-\beta)}{1-(K-1)e^{-2 \beta /\xi}}\right]^{\dfrac{1}{2\beta}}\,.
\label{eq:wc-beta}
}
Within the ULA, the critical disorder is given by 
\eq{
W_c^{\rm UL}(\xi,K)  = W_\ast(\beta_c,\xi,K)\,,
\label{eq:Wc-ULA}
}
where $\beta_c$ is fixed by the condition
\eq{
\partial_\beta W_\ast(\beta,\xi,K)\big\vert_{\beta=\beta_c} = 0\,,
\label{eq:betac-ULA}
}
and $\xi\ln\sqrt{K-1} <\beta_c<1/2$ for the solution to be valid.
Physically, this corresponds to picking the $\beta$ which minimises the solution to Eq.~\ref{eq:wc-beta} and thus yields a meaningful upper limit to the critical disorder. 

\begin{figure}[!b]
    \centering
    \includegraphics[width=\linewidth]{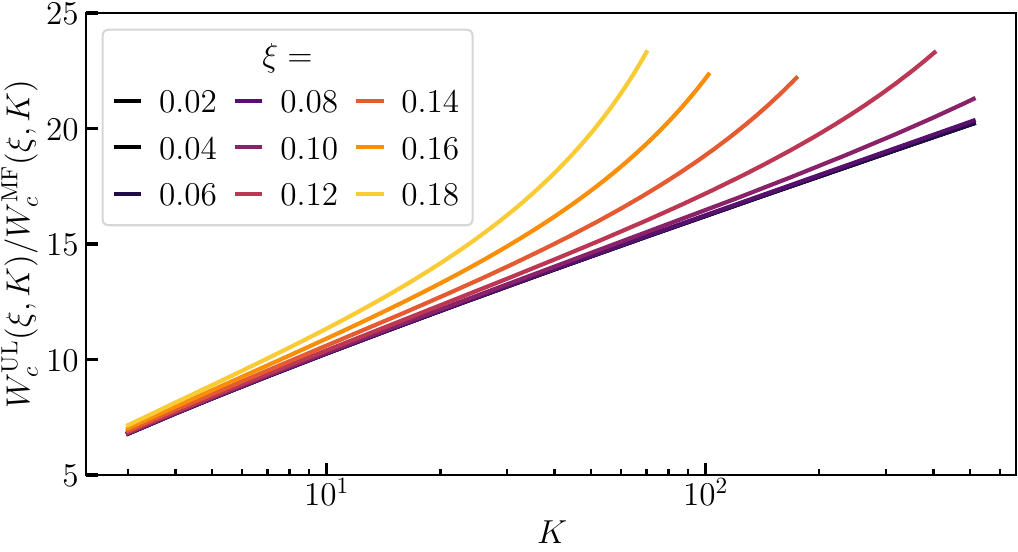}
    \caption{\new{The ratio $W_c^{\rm UL}(\xi,K)/W_c^{\rm MF}(\xi,K)$ as a function of $K$ for different 
    $\xi$ for the ExpRRG model. At very small $\xi$'s ($\lessapprox 0.1$),  $\ln K$ correction, typical to short-range Anderson model on tree-like graphs, is observed. On increasing $\xi$, at large $K$ values the correction grows faster than $\ln K$ manifesting a crucial interplay between $\xi$ and $K$. Above a certain $K$, there exists no $\beta_c>\xi \ln \sqrt{K-1}$ and therefore suggests the inapplicability of Eq.~\ref{eq:ula-condition}.
    }    }
    \label{fig:ula-mf-comparison}
\end{figure}

Obtaining a closed form expression 
for the solutions to Eq.~\ref{eq:Wc-ULA} and Eq.~\ref{eq:betac-ULA} turns out to be quite unwieldy. 
We therefore obtain solutions to them numerically.
Since our primary interest is in understanding the correction to the mean-field result in Eq.~\ref{eq:Wc-mf-loc}, we plot the ratio of $W_c^{\rm UL}(\xi,K)$ and the mean-field result which we denote as $W_c^{\rm MF}(\xi,K)$ as a function of $K$, in Fig.~\ref{fig:ula-mf-comparison}.
The results show that for $\xi\ll 1$, the correction is simply a multiplicative factor $\propto \ln K$, analogous to nearest-neighbour Anderson models on trees and RRGs.

However, as $\xi$ is tuned away from this limit
the correction  has a non-trivial dependence on $K$ and grows faster than $\ln K$ suggesting a complex interplay between the lengthscale $\xi$ and the connectivity $K$ of the graph. 
}

\bibliography{references}

\end{document}